%% file: main.tex
\documentclass[preprint]{elsarticle} 

\usepackage{amsfonts,mathtools,bm,amssymb}
\usepackage{subcaption}
\usepackage[inline]{enumitem}
\usepackage[dvipsnames]{xcolor}
\usepackage{amsthm,thmtools}
\usepackage{tikz}
\usetikzlibrary{shapes,arrows,fit,calc,positioning}
\usepackage{siunitx}
\usepackage{algorithm}
\usepackage{algpseudocodex}
\usepackage{multirow}
\mathtoolsset{showonlyrefs}

\graphicspath{{./Figures/}}

\declaretheorem{theorem}
\declaretheorem{corollary}
\declaretheorem{proposition}
\declaretheorem{lemma}

\declaretheorem[style=definition]{example}
\declaretheorem[style=remark]{remark}

\def\given{\,|\,}

\journal{Signal Processing}

\begin{document}

\begin{frontmatter}

\title{Decentralised possibilistic inference with applications to target tracking}

\author[1]{Jeremie Houssineau}
\ead{jeremie.houssineau@ntu.edu.sg}

\author[2]{Han Cai\corref{cor1}}
\ead{caihanspace@gmail.com}
\cortext[cor1]{Corresponding author}

\author[3]{Murat Uney}
\ead{M.Uney@ed.ac.uk}

\author[4]{Emmanuel Delande}
\ead{emmanuel.delande@cnes.fr}

\affiliation[1]{organization={Division of Mathematical Sciences, Nanyang Technological University},
            city={Singapore},
            postcode={637371}, 
            country={Singapore}}

\affiliation[2]{organization={School of Aerospace Engineering, Beijing Institute of Technology},
            city={Beijing},
            postcode={100081}, 
            country={China}}

\affiliation[3]{organization={Institute for Imaging, Data and Communications, University of Edinburgh},
            city={Edinburgh},
            postcode={EH9 3JL}, 
            country={UK}}

\affiliation[4]{organization={Centre National d'Etudes Spatiales},
            city={Toulouse},
            postcode={31400}, 
            country={France}}

\begin{abstract}
Fusing and sharing information from multiple sensors over a network is a challenging task, partly due to the absence of a foundational rule for fusing probability distributions that preserves the independence of sources. To address this, we propose a decentralised inference framework based on possibility theory. Unlike probabilistic approaches that rely on ad-hoc averaging, we derive a principled fusion rule that is proven to be asymptotically exact, meaning it recovers the posterior of the optimal centralised possibilistic approach. We apply this rule to the possibilistic Bernoulli filter, leveraging its hierarchical nature to jointly infer data association and state estimation, distinct from standard decentralised Kalman filtering. We demonstrate that the proposed approach maintains the independence of local posteriors during fusion and, even under necessary approximations to handle Gaussian mixtures, significantly outperforms probabilistic geometric and arithmetic average fusion baselines in terms of cardinality and localisation error.
\end{abstract}

\begin{keyword}
Target tracking \sep possibility theory \sep decentralised fusion
\end{keyword}

\end{frontmatter}

\section{Introduction}

Over the last decade, the increasing communication and computation capacity have made the view of multiple agents performing distributed inference particularly relevant in multi-sensor processing applications~\cite{Cetin2006}, in distributed estimation~\cite{Kar2008,Uney2011,Kar2013} in general and in target tracking~\cite{Liggins1997, Hall2017, Uney2013,Li2020on} in particular. There are challenges involved, however, such as the loss of precision that most approaches display when compared to centralised algorithms.

This work aims to highlight the capabilities in sensor fusion that can be leveraged when following an alternative version of Bayesian inference based on a particular formulation \cite{Houssineau2019, Houssineau2018parameter} of possibility theory \cite{Dubois2015}. Recently-derived target-tracking algorithms~\cite{Ristic2019, Houssineau2021linear, cai2022possibility} have shown that this approach allows for a greater uncertainty about some crucial aspects of target tracking when compared to their probabilistic analogues. Yet, their potential for sensor fusion is unrelated to these advantages and pertains instead to the fundamental nature of possibility theory as a representation of information rather than randomness. Indeed, as will be demonstrated in this work, information can be naturally shared, and combined~\cite{dubois1994possibility, dubois1999merging} in this context whereas these operations are less straightforward in probability theory and require additional principles and approximations.

In order to demonstrate the generality of the proposed approach to decentralised fusion, we apply it to a target-tracking problem \cite{BarShalom1990, Mahler2007} where an object of interest, the \emph{target}, is observed under a partial, noisy and corrupted observation process:
\begin{enumerate*}[label=\roman*)]
\item \emph{partial} because the state of the target is not fully observed and because its detection can fail altogether, an event we refer to as a detection failure,
\item \emph{noisy} because the observed components of the target's state are subject to observation errors, and
\item \emph{corrupted} because observations which do not originate from the target are also collected; we refer to these as false alarms.
\end{enumerate*}
The target tracking problem is made challenging by the absence of information on the data association, i.e.\ the observations do not convey any direct information regarding which of them are false alarms or originated from targets, hence making the problem combinatorially complex for the multitude of the association-related hypotheses. This difficulty is exacerbated by the uncertainties in the location of the target when it first appears and the time at which it disappears; these are often modelled by --- borrowing from the population statistics jargon --- a birth/death process. Solutions to this problem are often referred to as \emph{Bernoulli filters} \cite{ristic2013tutorial}, with ``Bernoulli'' referring to the presence/absence of the target due to the birth/death process. A possibilistic Bernoulli filter has been introduced in \cite{Ristic2019}. Distributed fusion of target tracking has attracted a lot of attention in the last few years, considering for instance sensors with limited field of view \cite{Uney2016,Yi2020,reifler2024improving}.

In probabilistic track-to-track fusion, geometric averaging is closely related to Chernoff (exponential-mixture) fusion, which is widely used as a conservative rule under unknown cross-correlations. Efficient sigma-point approximations have been proposed for Chernoff fusion of Gaussian mixtures \citep{gunay2017chernoff}. Alternative pooling rules beyond arithmetic/geometric means have also been studied for mixture densities, e.g. harmonic-mean density pooling \citep{sharma2026pooling}.

One of the objectives of this work is to show how the proposed approach provides a principled sensor fusion framework that can be easily applied to a complex inference algorithm, such as the possibilistic Bernoulli filter. Our interest in Bernoulli filters is underpinned by their hierarchical nature that infers data association and state estimation jointly in contrast to modern decentralised Kalman filtering techniques (e.g.\ \cite{talebi2019distributed}), which do not address the data association problem. The main contributions of this work are as follows:
\begin{enumerate}[wide,nosep]
    \item A decentralised possibilistic Bayesian inference method is introduced and proved to be algebraically equivalent to the centralised solution.
    \item The proposed approach is detailed in the case of a Gaussian mixture implementation of the possibilistic Bernoulli filter, highlighting its applicability to inference problems beyond the linear-Gaussian setting of the Kalman filter. The Gaussian mixture is a hurdle for technique like geometric averaging, which lack closed-form solutions for the power of a mixture.
    \item The performance of the proposed decentralised possibilistic Bernoulli filter is shown to largely improve existing approaches \cite{guldogan2014consensus,li2019distributed}, verifying the claim that the performance in the centralised case can be closely matched even with the necessary approximations and practical considerations (such as a small number of communication steps on the sensor network).
\end{enumerate}

In subsequent work \cite{cai2024robust}, we extend the present framework to labelled multi-Bernoulli multi-target tracking. In contrast, this paper makes the foundational contribution: it develops a distributed possibilistic fusion rule and proves that, under stated conditions, repeated decentralised fusion recovers the corresponding centralised (``oracle'') result. The tracking study included here is intentionally controlled to validate the fusion behaviour and its information-loss properties, whereas \cite{cai2024robust} addresses additional multi-target modelling and algorithmic layers beyond the fusion operator.

The structure of this article is as follows: An introduction to possibility theory is given in Section~\ref{sec:possibilityTheory}. The main result of the article is stated and proved in Section~\ref{sec:decentralisedBayesianInference}. In order to illustrate this result for a non-trivial inference problem, we review the possibilistic Bernoulli filter in Section~\ref{sec:BernoulliFilter}, detail the corresponding sensor fusion methodology in Section~\ref{sec:sensorFusion}, and assess its performance on simulated data in Section~\ref{sec:simulations}. The article concludes in Section~\ref{sec:conclusion}.

\section{Review of possibilistic inference}
\label{sec:possibilityTheory}

Possibility theory provides a way to represent \emph{imprecise knowledge} about an unknown (but fixed) quantity of interest \cite{Dubois2015,deBaets1999}. As in probability theory, we introduce a sample space $\Omega$. However, instead of placing a probability distribution on $\Omega$, we assume there is a true state of nature $\omega^* \in \Omega$ and we seek to infer whatever features of $\omega^*$ matter for the problem at hand.

An analogue of a random variable is then defined: an \emph{uncertain variable} $\bm{x}$ is a mapping from $\Omega$ to a set $S$. If $\omega$ is the true state of nature, then the corresponding true value is $x=\bm{x}(\omega)$.

The information available about $\bm{x}$ is represented by a non-negative function $f_{\bm{x}}$ such that $\sup_{x \in S} f_{\bm{x}}(x)=1$. This function is called a \emph{possibility function} (p.f.). Under this representation, the event $\bm{x}\in B$ (for $B\subseteq S$) is assigned the \emph{credibility}
\[
\sup_{x\in B} f_{\bm{x}}(x).
\]
Informally, credibility is a degree of belief. More formally, it can be viewed as the largest \emph{subjective} probability an agent is willing to assign to the event. The probability is subjective because $\bm{x}$ is not modelled as a random variable, so there is no underlying ``true'' probability of the event.

A key aspect of possibility theory is that p.f.s are inherently agent-dependent: different agents may have different information about the same unknown quantity, and therefore different p.f.s. The least informative belief is the p.f.\ that equals $1$ everywhere on $S$. We denote it by $\bm{1}$. It expresses complete ignorance and, when $S$ is unbounded, it has no direct analogue as a proper probability density. More informative p.f.s satisfy $f_{\bm{x}}(x)<1$ for some $x\in S$.

An agent may deliberately discard information by replacing $f_{\bm{x}}$ with another p.f.\ $f'_{\bm{x}}$ such that $f_{\bm{x}}\le f'_{\bm{x}}$, i.e.\ $f_{\bm{x}}(x)\le f'_{\bm{x}}(x)$ for all $x\in S$. This may be useful for analytical or computational simplicity. In terms of the ``maximum subjective probability'' interpretation, $f'_{\bm{x}}$ yields a larger (or equal) maximum probability for every event, hence a wider set of probabilities consistent with the agent's belief. In the extreme case $f'_{\bm{x}}=\bm{1}$, any event can be assigned any probability between $0$ and $1$, which corresponds to having no information.

A central difference between probability density functions (p.d.f.s) and p.f.s is that p.f.s are not densities. In particular, $f_{\bm{x}}(x)$ can be interpreted directly as the credibility of the event $\bm{x}=x$. This leads to a different change-of-variables rule. For any mapping $T$ on $S$, the transformed uncertain variable $\bm{x}'=T(\bm{x})$ is described by
\[
f_{\bm{x}'}(x')=\sup_{x\in T^{-1}(x')} f_{\bm{x}}(x),
\]
for any $x'$ in the image of $T$. Here $T^{-1}(x')$ denotes the (possibly set-valued) pre-image of $x'$, and we take $\sup\emptyset=0$ when no $x$ maps to $x'$. If $T$ is bijective, this simplifies to $f_{\bm{x}'}(x')=f_{\bm{x}}(T^{-1}(x'))$, with no Jacobian term (in contrast to the probabilistic case).

\subsection{Independence and marginal possibilities}

If another uncertain variable $\bm{y}$ is defined in a set $S'$, then the joint information about $\bm{x}$ and $\bm{y}$ can be modelled by a p.f.\ $f_{\bm{x},\bm{y}}$ on $S \times S'$. The marginal p.f.\ of $\bm{x}$ induced by $f_{\bm{x},\bm{y}}$ is denoted by $f_{\bm{x}}$ and given by
\begin{equation}
\label{eq:marginalisation}
f_{\bm{x}}(x) = \sup_{y \in S'} f_{\bm{x},\bm{y}}(x,y), \qquad x \in S.
\end{equation}

The uncertain variables $\bm{x}$ and $\bm{y}$ are said to be \emph{independently described} by $f_{\bm{x},\bm{y}}$ if there exist p.f.s $f_{\bm{x}}$ and $f_{\bm{y}}$ such that
\begin{equation}
f_{\bm{x},\bm{y}}(x,y) = f_{\bm{x}}(x) f_{\bm{y}}(y), \qquad \forall (x,y) \in S \times S'.
\label{eqn:Independence}
\end{equation}
This definition of independence follows so-called \emph{numerical} possibility theory, although other notions of independence exist in possibility theory \cite{dubois2000possibility}.

Numerical possibility theory has strong connections with probability theory, as will be seen in Section~\ref{sec:posteriorPosssibility} where the analogues of Bayes theorem and Gaussian distributions will be defined; however, despite these connections, the considered notion of independence differs fundamentally from the probabilistic one as it is not an intrinsic property of the considered uncertain variables, but rather a statement about the absence of a relation between the information we hold about $\bm{x}$ and the one we hold about $\bm{y}$. In particular, uncertain variables can be \emph{assigned} joint p.f.s that exhibit the independence property in \eqref{eqn:Independence}, e.g., by bringing p.f.s to a given power, an operation generally referred to as discounting. As opposed to p.d.f.s, p.f.s are closed under exponentiation, i.e.\ $f_{\bm{x}}^{w}$ is still a non-negative function with a supremum that equals to $1$ for any $w \in [0,1]$; in particular, $f_{\bm{x}}^0$ is equal to $\bm{1}$.

The next step is to introduce ways of combining information; this is achieved through Bayes' theorem in the following section.

\subsection{Posterior possibility and conjugate priors}
\label{sec:posteriorPosssibility}

Assuming that the event $\bm{y} = y$ has positive credibility, i.e.\ $f_{\bm{y}}(y) > 0$, the analogue of Bayes theorem for p.f.s was introduced in~\cite{deBaets1999} as
\begin{align}
\label{eq:BayesRule}
f_{\bm{x}|\bm{y}}(x \given y) & = \dfrac{f_{\bm{x},\bm{y}}(x,y)}{f_{\bm{y}}(y)} \nonumber \\
=& \dfrac{f_{\bm{y}|\bm{x}}(y \given x) f_{\bm{x}}(x)}{\sup_{x' \in S} f_{\bm{y}|\bm{x}}(y \given x') f_{\bm{x}}(x')},
\end{align}
for any $x \in S$. 

The concept of a \emph{conjugate prior} translates directly to this form of posterior inference, e.g., if $S = \mathbb{R}^d$ and $S' = \mathbb{R}^{d'}$ and if both the \emph{prior} $f_{\bm{x}}$ and the \emph{likelihood} $f_{\bm{y}|\bm{x}}(y \given \cdot)$ take the form of a Gaussian (quadratic exponential) p.f., i.e.
\[
f_{\bm{x}}(x) = \overline{\mathrm{N}}(x; \mu, P) = \exp\Big( -\dfrac{1}{2} (x - \mu)^{\top} P^{-1} (x - \mu) \Big)
\]
for some $\mu \in S$ and some $d \times d$ positive definite matrix $P$, and $f_{\bm{y}|\bm{x}}(y \given x) = \overline{\mathrm{N}}(y; H x, R)$ for
some $d' \times d$ matrix $H$ and some $d' \times d'$ positive definite matrix $R$, then the \emph{posterior} p.f.\ $f_{\bm{x}|\bm{y}}(\cdot \given y)$ is also Gaussian \cite{Houssineau2018}. Every probabilistic conjugate prior has a possibilistic analogue, up to differences in the set of parameters for which a given conjugate prior family is well-defined, due to the fact that p.f.s need to be bounded whilst p.d.f.s need to  integrate to unity~\cite{Houssineau2019}.

Using more advanced tools, a version of the Bayes theorem with a possibilistic prior and a probabilistic likelihood is derived in \cite{Houssineau2018parameter} and proved to be
\begin{equation}
\label{eq:mixedBayes}
f_{\bm{x}|Y}(x \given y) = \dfrac{p(y \given x)  f_{\bm{x}}(x)}{\sup_{x' \in S} p(y \given x')  f_{\bm{x}}(x')},
\end{equation}
where $y$ is a realisation of a random variable $Y$ whose probability density function given $\bm{x} = x \in S$ is $p(\cdot \given x)$. This alternative version is useful as it is common for the observation process to involve randomness even when the dynamics of the target do not.

\section{Managing information with possibility theory}

We start with a result about the general fusion of two sources of information, where ``general'' refers to the fact that the considered sources of information are not necessarily independent. The proofs of the results in this section and the next are found in Appendix~A. 
The following proposition highlights that conditioning on events of the form $\bm{x} = \bm{z}$ is straightforward in possibility theory, contrary to probabilistic conditioning (see, e.g., \cite[Chapter 15.7]{Jaynes2003}).

\begin{proposition}
\label{prop:generalFusion}
For uncertain variables $\bm{x}$ and $\bm{z}$ on the same set $S$, suppose that some p.f.\ $f_{\bm{x},\bm{z}}$ is their joint descriptor. Then, the posterior p.f.\ describing $\bm{x}$ (equivalently $\bm{z}$) given that $\bm{x}$ and $\bm{z}$ represent the same unknown quantity, i.e.\ $\bm{x} = \bm{z}$, is given by
\[
f_{\bm{x}}(x \given \bm{x} = \bm{z}) = \dfrac{f_{\bm{x},\bm{z}}(x,x)}{\sup_{x' \in S}f_{\bm{x},\bm{z}}(x',x')}.
\]
\end{proposition}

Alternative fusion rules in the context of possibility theory are studied in \cite{destercke2008possibilistic}. To simplify the notation, we will often denote a fused p.f.\ such as $f_{\bm{x}}(\cdot \given \bm{x}  = \bm{z})$ by $\hat{f}$. The following corollary is a straightforward generalisation of Proposition~\ref{prop:generalFusion} for an arbitrary number of independent sources of information.

\begin{corollary}
\label{cor:fusionNsources}
If $\{f_i\}_{i=1}^n$ is a collection of $n > 1$ independent p.f.s on $S$ representing the same unknown quantity, then the corresponding fused p.f.\ is
\begin{equation}
\label{eq:fusensources}
\hat{f}(x) = \dfrac{\prod_{i=1}^n f_i(x)}{\sup_{x' \in S}\prod_{i=1}^n f_i(x')}.
\end{equation}
\end{corollary}

We now show how to split a single source of information into two independent pieces with no loss of information, i.e.\ these two pieces of information can be fused back into the original one. This is crucial for decentralised fusion where information is sequentially shared~/~fused. To simplify the presentation, when two uncertain variables $\bm{x}$ and $\bm{z}$ are independently described by some $f_{\bm{x}}$ and $f_{\bm{z}}$, we will just say that ``$f_{\bm{x}}$ and $f_{\bm{z}}$ are independent''; similarly, when it is known that $\bm{x} = \bm{z}$, we will just say ``$f_{\bm{x}}$ and $f_{\bm{z}}$ represent the same unknown quantity''.

\begin{proposition}
\label{prop:splitAndMerge}
The information captured in the p.f.\ $f$ is equivalent to a combination of that in the two independent p.f.s $f^w$ and $f^{1-w}$, for any $w \in [0,1]$, when it holds that $f^w$ and $f^{1-w}$ represent the same unknown quantity.
\end{proposition}

\begin{figure}
\centering
 \begin{subfigure}[b]{0.48\textwidth}
        \centering
        \includegraphics[width=\textwidth]{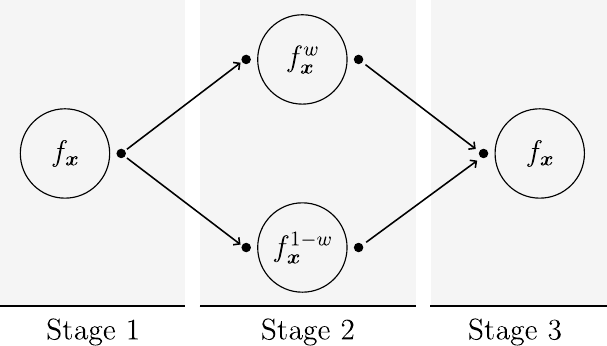}
        \caption{Recovery of the prior p.f.\ $f_{\bm{x}}$ by merging after splitting.}
        \label{fig:SplitMerge}
    \end{subfigure}
    \begin{subfigure}[b]{0.48\textwidth}
        \centering
        \includegraphics[width=\textwidth]{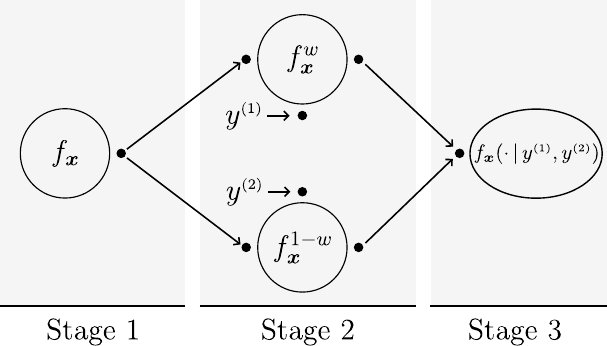}
        \caption{Recovery of the posterior $f_{\bm{x}}(\cdot \given y^{(1)}, y^{(2)})$ by merging local posteriors based on split priors.}
        \label{fig:SplitMergePosterior}
    \end{subfigure}
\caption{3-stage process to split and merge the information in a p.f., with and without additional conditionally-independent observations $y^{(1)}$ and $y^{(2)}$.}
\label{fig:stages}
\end{figure}

In practice, Proposition~\ref{prop:splitAndMerge} implies that a p.f.\ $f$ can be perfectly recovered after splitting it into $f^{w}$ and $f^{1-w}$, regardless of the weight $w \in [0,1]$. This means that there is no need to identify further principles to optimise with respect to $w$. Graphically, these steps are illustrated in \figurename~\ref{fig:stages}\subref{fig:SplitMerge}, where each stage follows from principled operations on p.f.s. We are interested in the case where additional (conditionally independent) information is acquired at each node in Stage~2 via the observations $y^{(1)}$ and $y^{(2)}$, which will retain the independence necessary for the fusion in Stage~3 (\figurename~\ref{fig:stages}\subref{fig:SplitMergePosterior}). The fact that the result of this fusion operation is the posterior p.f.\ $f_{\bm{x}}(\cdot \given y^{(1)}, y^{(2)})$ will be shown in Section~\ref{sec:decentralisedBayesianInference}.

This result already hints at the capabilities of possibility theory in terms of distributed inference: One can share information across a sensor network, update it with new observations and then recombine all this information in a natural way, without losses due to approximations or errors inducing spurious information, as depicted in \figurename~\ref{fig:stages}.

\section{Decentralised possibilistic Bayesian inference}
\label{sec:decentralisedBayesianInference}

In this section, we introduce a decentralised possibilistic Bayesian inference scheme and prove that it is algebraically equivalent to the centralised solution. We consider a prior p.f.\ $f$ on a given set $S$ and model the evolution of the underlying state by a (possibilistic) Markov transition $g(\cdot \given x)$, which is a p.f.\ on $S$ for any $x \in S$. In this context, if $n$ observations $y_1,\dots,y_n$ are received and if these observations are described by a p.f.\ of the form $h(y_1,\dots,y_n \given x) = \prod_{i=1}^n h^{(i)}(y_i \given x)$, for any $x \in S$, then the standard (centralised) approach is to consider the posterior p.f.\
\begin{equation}
\label{eq:StandardBayes}
f(x \given y_1,\dots,y_n) \propto \bigg[ \prod_{i=1}^n h^{(i)}(y_i \given x) \bigg] \sup_{x' \in S} g(x \given x') f(x').    
\end{equation}
This posterior p.f.\ is what we are trying to recover in the decentralised case.

We now consider $n$ sensor nodes $\mathrm{S}_i$ with $i \in \mathcal{V} = \{1,\dots,n\}$ and assume that the sensor network is modelled by an undirected connected graph $\mathcal{G} = (\mathcal{V},\mathcal{E})$ where $\mathcal{E}$ is the set of edges over which the nodes can communicate. Each node collects its independent observations of the same underlying variable. We refer to this configuration as decentralised fusion. We will use the superscript $\cdot^{(i)}$ to refer to quantities that are specific to sensor node $\mathrm{S}_i$. We denote by $\mathcal{N}_i = \{ j \in \mathcal{V} : \{i,j\} \in \mathcal{E} \}$ the set of neighbours of $i \in \mathcal{V}$. We assume that the total number of nodes in the network is known by every node; this can indeed be obtained in a decentralised manner via standard techniques \cite{Jesus2014}. The objective in this section is to show that the posterior p.f.\ $f(\cdot \given y_1,\dots,y_n)$ defined in \eqref{eq:StandardBayes} can be recovered by alternating between communication on the network and fusion of the received information at each node using~\eqref{eq:fusensources}.

One step of the decentralised algorithm is specified as follows: let us first follow Proposition~\ref{prop:splitAndMerge} and share the prior information across the network after splitting into $f^{(i)} = f^{\omega_i}$ at node $i \in \mathcal{V}$ for some collection of non-negative weights $\{\omega_i\}_{i\in\mathcal{V}}$. Performing prediction at each node with the Markov transition $g$ would cause a large overlap between each local p.f.\: if all nodes were to be connected to a central node $\mathrm{C}$ and fused, there would be no way of recovering the predicted p.f.\ $\sup_{x \in S} g(\cdot \given x) f(x)$, as shown in the following example.

\begin{example}
Consider the two sensor case, say $\mathcal{V} = \{1,2\}$. Let $f$ be the standard Gaussian p.f.\ $\overline{\mathrm{N}}(0,1)$, so that $f^{(i)} = \overline{\mathrm{N}}(0,1/\omega_i)$ for $i = 1,2$. We assume that $\omega_1 + \omega_2 = 1$ so that fusing the priors back together indeed yields $\overline{\mathrm{N}}(0,1)$. We also assume that $\omega_1 \in (0,1)$ to avoid trivial situations. If we predict at each node, with the Markov transition $g(x\given x') = \overline{\mathrm{N}}(x',\tau^{-1})$ with precision $\tau$, then the predicted p.f.\ at node $i$ will be $\overline{\mathrm{N}}(0,1/\tau + 1/\omega_i)$. Fusing the information from all nodes will yield the p.f.\ $\overline{\mathrm{N}}(0, \hat{\tau}^{-1})$ with precision $\hat{\tau} = \frac{\tau\omega_1}{\tau + \omega_1} + \frac{\tau\omega_2}{\tau + \omega_2} > \frac{\tau}{\tau+1}$. The centralised predictive p.f.\ is $\overline{\mathrm{N}}(0, (\tau+1)/\tau)$, so we will always be overly optimistic about the predicted precision when fusing nodes where prediction has been carried out without discounting the Markov transition.
\end{example}

Instead, we share the information about the evolution of the state by considering the Markov transition $g^{\varpi_i}(\cdot \given x)$ at $\mathrm{S}_i$, for some collection of non-negative weights $\{\varpi_i\}_{i\in\mathcal{V}}$. Finally, we assume that the observations are local, i.e. $y_i$ is the only observation at $\mathrm{S}_i$. As a result, the local posterior p.f.\ at node $i$ is computed as $f^{(i)}(x \given y_i) \propto h^{(i)}(y_i \given x) \sup_{x' \in S} g^{\varpi_i}(x \given x') f^{\omega_i}(x')$.

The method we propose is reminiscent of iterative message-passing approaches to decentralised inference~\cite{Cetin2006}: we assume that each node $j \in \mathcal{V}$ communicates with its neighbours and combines the information it receives by using a weight matrix $\Gamma = (\Gamma_{i,j})_{i,j \in \mathcal{V}}$ such that $\Gamma_{i,j} \geq 0$ for all $i,j \in \mathcal{V}$ and such that $\Gamma_{i,j}=0$ when $i\notin \mathcal{N}_j$.
The corresponding fusion of information is carried out via the recursion
\begin{align}
\label{eqn:LocalUpdate}
f^{(j,l)}(x) & \propto {\Big( f^{(j,l-1)}(x)\Big)^{\Gamma_{j,j}} \prod_{i \in \mathcal{N}_j } \Big( f^{(i,l-1)}(x)\Big)^{\Gamma_{i,j}} },\\
& = \prod_{i \in \mathcal{V} } \Big( f^{(i,l-1)}(x)\Big)^{\Gamma_{i,j}} \quad  l=1,\dots,L \nonumber
\end{align}
with the initial condition being selected as $f^{(i,0)} = f^{(i)}( \cdot \given y_i)$. The following result shows that, crucially, such an approach preserves the independence of the local fused posterior p.f.s.

\begin{corollary}
\label{cor:preservesIndependence}
Assume that $\Gamma$ is right stochastic, i.e., $\sum_{j \in \mathcal{V}} \Gamma_{i,j} = 1$ all $i \in \mathcal{V}$, and that the p.f.s $f^{(i)}( \cdot \given y_i)$, $i \in \mathcal{V}$, are mutually independent, then the recursion \eqref{eqn:LocalUpdate} preserves the mutual independence of the p.f.s $f^{(j,l)}$, $j \in \mathcal{V}$, for all steps $l=1,\dots,L$.
\end{corollary}

To simplify the study of the asymptotic properties of the proposed fusion algorithm, the locally fused result in \eqref{eqn:LocalUpdate} can be related to the $l$th power of the matrix $\Gamma^l$ by
$
f^{(j,l)}(x) = \prod_{i \in \mathcal{V} } \big( f^{(i)}(x \given y_i)\big)^{\Gamma^l_{i,j}}$, for $l=1,\dots,L$. The limiting case is captured by the matrix $\Gamma^*$ defined via the limit $\Gamma^{l} \xrightarrow{l\to\infty} \Gamma^*$. In turn, the $(i,j)$-th entry of $\Gamma^*$ is denoted by $\Gamma^*_{i,j}$. 

We now investigate the conditions under which the posterior p.f.\ $f(\cdot \given y_1,\dots,y_n)$ can be recovered from the asymptotic posterior $f^{(i,\infty)}$ at any node $i \in \mathcal{V}$. Since it is particularly convenient to have $1/n$-th of the information at each node, we consider the relation $f^{(i,\infty)}(x) = f(x \given y_1,\dots,y_n)^{1/n}$.
Algorithms verifying this relation are said to be ``asymptotically exact''. We consider the conditions:
\begin{enumerate}[label={A\arabic*},nosep]
    \item \label{A1} It holds that $\omega_i = \varpi_i$ for all $i \in \mathcal{V}$ and that $\sum_{i \in \mathcal{V}} \omega_i = 1$
    \item \label{A2} It holds that $\Gamma^*_{i,j} = 1/n$ for all $i,j \in \mathcal{V}$
\end{enumerate}

\begin{theorem}
\label{thm:decentralisedBayesianInference}
For any number $n$ of sensors, Assumptions~\ref{A1} and \ref{A2} are necessary and sufficient for the proposed decentralised Bayesian inference algorithm to be asymptotically exact.
\end{theorem}

Some crucial aspects of Theorem~\ref{thm:decentralisedBayesianInference} should be noted: Firstly, it makes no assumptions on the nature of the set $S$ on which the fusion is performed. In particular, it can be applied for variables in hierarchical problems such as the ones appearing in multi-target tracking, where both the number of targets and their respective state must be inferred. We illustrate this aspect in Section~\ref{sec:sensorFusion} where fusion is performed in a case where the existence of a target is uncertain; we then show in Section~\ref{sec:simulations} that Theorem~\ref{thm:decentralisedBayesianInference} can be verified in practice for such problems despite the use of approximations. Secondly, the assumptions are proved to be necessary, that is, they cannot be relaxed without considering models with special properties, such as those involving indicator functions which are unaffected by exponentiation.

We now consider an immediate corollary to Theorem~\ref{thm:decentralisedBayesianInference} which considers the case where $\mathcal{G}$ is complete, i.e.\ there is an edge between every pair of nodes, with the following modified assumption:
\begin{enumerate}[label=A\arabic*,resume,nosep]
    \item \label{A3} It holds that $\mathcal{G}$ is complete and that $\Gamma_{i,j} = 1/n$ for all $i,j \in \mathcal{V}$.
\end{enumerate}

\begin{corollary}
\label{thm:centralisedBayesianInference}
For any number $n$ of sensors, Assumptions~\ref{A1} and \ref{A3} are necessary and sufficient for the proposed decentralised Bayesian inference algorithm to be exact.
\end{corollary}

Corollary~\ref{thm:centralisedBayesianInference} can be proved by noticing that under Assumption A3, it holds that $\Gamma = \Gamma^*$, i.e., the asymptotic regime can be achieved in a single iteration.

\begin{remark}
The results of Theorem~\ref{thm:decentralisedBayesianInference} and Corollary~\ref{thm:centralisedBayesianInference} still hold if we replace the p.f.\ $h(y_1,\dots,y_n \given x)$ by a probability distribution function (p.d.f.)\ of the form $p(y_1,\dots,y_n \given x) = \prod_{i=1}^n p^{(i)}(y_i \given x)$. Indeed, the corresponding posterior, obtained through the form of Bayes theorem defined in \eqref{eq:mixedBayes}, remains a p.f.\ and can therefore be discounted and/or fused similarly. In fact, the likelihood can also be a more complex object that involves both p.f.s and p.d.f.s with no effect on the result. An example of such a likelihood is provided in Section~\ref{sec:BernoulliFilter}.
\end{remark}

\begin{remark}
In the linear-Gaussian case, one can use the possibilistic version of the Kalman filter \cite{Houssineau2018} to perform Bayesian inference. The possibilistic Kalman filter has the same expected value and variance as the standard Kalman filter and, in the centralised case, the proposed approach reduces to the optimal fusion rule in the independent case~\cite{Kim1994}.
\end{remark}

Finally, we consider the case where node $i^* \in \mathcal{V}$ plays the role of a central node to which all the sensors are connected and which is the only one performing fusion. We refer to this scheme as centralised fusion. Decentralised fusion with $\Gamma$ such that $\Gamma_{i,j}$ equals $1$ if $j = i^*$ and $0$ otherwise is equivalent to centralised fusion; a minor modification in the proof of Theorem~\ref{thm:decentralisedBayesianInference} allows to show that centralised fusion can also be performed exactly under Assumption~\ref{A1}, meaning that $f^{(i^*,\infty)} = f(\cdot \given y_1,\dots,y_n)$ in this case.

\section{The possibilistic Bernoulli filter}
\label{sec:BernoulliFilter}

\subsection{General recursion}

\subsubsection{Independent observations}

In this section, we present a modified version of the equations of the possibilistic Bernoulli filter \cite{Ristic2019} and the underlying notion of Bernoulli uncertain finite set. The latter can be introduced as an uncertain variable $\bm{X}$ on the set $\mathsf{X} = \{\emptyset\} \cup \{ \{x\} : x \in S \}$, which is the set of finite subsets of a given space $S \subseteq \mathbb{R}^d$ with no more than one element. The uncertain finite set $\bm{X}$ can be described by a p.f.\ $F$ on $\mathsf{X}$, and we will adopt the convention
\begin{equation}
\label{eq:BernoulliPossibilityFunction}
F(X) = \begin{cases*}
\beta & if $X = \emptyset$ \\
\alpha f(x) & if $X = \{x\}$ with $x \in S$.
\end{cases*}
\end{equation}
The fact that $F$ is a p.f.\ implies that $\alpha$ and $\beta$ are non-negative, $\max\{\alpha, \beta\} = 1$, and $f$ is a p.f.\ on $S$. The possibilistic Bernoulli filter is a recursion for a p.f.\ of the same form as $F$ describing the possibility of existence $\alpha$ of a target in $S$ and the information about its state via $f$. We follow the notations of Section~\ref{sec:decentralisedBayesianInference} to describe a target's dynamics and observation when it exists. We model the changes in cardinality by a transition matrix $(\tau_{ij})_{i,j \in \{0,1\}}$, where $\tau_{ij}$ corresponds to transitioning from cardinality $i\in \{0,1\}$ to cardinality $j \in \{0,1\}$. For instance, $\tau_{01}$ is the possibility of birth. The spatial information about birth is described by a p.f.\ $f_{\mathrm{b}}$ on $S$. Denoting by $F_{k-1}$ the Bernoulli p.f.\ describing the state and existence of the target at time $k-1$, given the observations up to time $k-1$, and following the convention of \eqref{eq:BernoulliPossibilityFunction}, we can now express $F_{k|k-1}$, the predicted Bernoulli p.f.\ at $k$, as
\begin{equation}
\label{eq:BernoulliPrediction}
F_{k|k-1}(X) = \begin{cases*}
\beta_{k|k-1} & if $X = \emptyset$ \\
\alpha_{k|k-1} f_{k|k-1}(x) & if $X = \{x\}$ with $x \in S$,
\end{cases*}
\end{equation}
with $\beta_{k|k-1} = \max\{ \beta_{k-1} \tau_{00}, \alpha_{k-1} \tau_{10} \}$ and $\alpha_{k|k-1} = \max\{ \beta_{k-1} \tau_{01}, \alpha_{k-1} \tau_{11} \}$, and with
\[
f_{k|k-1}(x)  = \alpha_{k|k-1}^{-1}\max\big\{ \beta_{k-1} \tau_{01} f_{\mathrm{b}}(x), \alpha_{k-1} \tau_{11} \tilde{f}(x) \big\},
\]
where $\tilde{f}(x) = \sup_{x' \in S} g(x \given x') f_{k-1}(x')$ is the predicted p.f.\ describing the state of the target given that it exists at both times $k-1$ and~$k$.

As opposed to the approach in \cite{Ristic2019}, we consider a partially-probabilistic observation model where the detection and the data association are possibilistic and where the generation of false alarms and of the observation under the assumption of detection are probabilistic. We assume as is usual that at most one observation originates from the target. This yields a likelihood $\ell(Y_k \given X)$ for the observation set $Y_k$ at time $k$ characterised by $\ell(Y_k \given \emptyset) = \kappa(Y_k)$ and
\begin{equation}
\label{eq:generalLikelihood}
\ell(Y_k \given \{x\}) = \max\big\{ \beta_{\mathrm{d}} \kappa(Y_k), \max_{y \in Y_k} \alpha_{\mathrm{d}} p(y \given x) \kappa(Y_k \setminus \{y\}) \big\},
\end{equation}
where $\kappa$ is the density of the random finite set characterising the false alarms, where $\alpha_{\mathrm{d}}$ and $\beta_{\mathrm{d}}$ are respectively the possibilities of detection and non-detection, and where $p(\cdot \given x)$ is the p.d.f.\ of the observation given that the target is detected at state $x \in S$. The likelihood $\ell( \cdot \given X)$ is neither a p.f.\ nor a p.d.f., and its derivation, which can be found in Appendix~B, requires more advanced notions. The introduction of the likelihood $\ell( \cdot \given X)$ is important as it allows to model false alarms and observation noise as random phenomena, which is well accepted in the literature, while modelling the more subjective aspects of the problem, such as prior information and the target's dynamics, with p.f.s. The availability and tractability of such a model also hints at the potential for introducing more sophisticated mixed models in the future.

Since the posterior ultimately depends only on $\kappa$ and $p(y \given \cdot)$ via ratios of the form $p(y \given x) \kappa(Y \setminus \{y\}) / \kappa(Y)$, the form of the likelihood is similar to the one in \cite{Ristic2019} and the updated p.f.\ $F_k$ can be found in the same way: Following the same convention as before, we obtain that $\beta_k \propto \ell(Y_k \given \emptyset)\beta_{k|k-1} = \kappa(Y_k) \beta_{k|k-1}$ and $\alpha_k \propto \alpha_{k|k-1} \sup_{x \in S} \ell(Y_k \given \{x\}) f_{k|k-1}(x)$ with $\max\{ \alpha_k, \beta_k \} = 1$, and that
\[
f_k(x) = \dfrac{\ell(Y_k \given \{x\}) f_{k|k-1}(x)}{\sup_{x \in S} \ell(Y_k \given \{x\}) f_{k|k-1}(x)}.
\]

\subsubsection{Partially-unknown dependence between sensors via a hybrid likelihood}
\label{sec:partialDependenceHybrid}

So far, we have assumed that, conditional on the target state, the observations collected at the different nodes are independent so that the (multi-sensor) likelihood factorises across sensors. This section shows how to retain the same distributed recursion when part of the observation mechanism induces dependence across sensors, by isolating the dependent component and replacing it by an independent (but conservative) surrogate.

Let $x\in S$ denote the (single-target) state at time $k$ and let $Y_k^{(i)}$ denote the observation set collected by node $\mathrm{S}_i$, $i\in \mathcal{V} = \{1,\dots,n\}$. We introduce an intermediate variable $\tilde{z}_k = [ z_k^{(1)},\dots,z_k^{(n)} ]^{\top} \in (S')^n$, intended to capture the correlated part of the sensing mechanism (e.g.\ an unknown object attribute or a common propagation effect). We model the mapping $x \mapsto \tilde z_k$ possibilistically by a p.f.\ $h_1(\tilde{z}_k\given x)$. Conditional on $z_k^{(i)}$, the remaining part of the observation mechanism at node $i$ is assumed probabilistic and independent across sensors. Concretely, we assume that the likelihood term that appears in the Bernoulli update at node $i$ can be written in the form $\ell_2^{(i)}(Y_k^{(i)}\given z_k^{(i)})$, where $\ell_2^{(i)}(\cdot\given z)$ is the usual single-sensor Bernoulli-set likelihood (false alarms, detection, and measurement noise) with the p.d.f.\ part now parameterised by $z$ rather than directly by $x$. The resulting hybrid multi-sensor likelihood at an observation $\tilde{Y}_k = (Y_k^{(i)})_{i\in \mathcal{V}}$ is then defined as
\begin{equation}
\label{eq:hybrid_multisensor_likelihood}
\ell(\tilde{Y}_k \given \{x\}) = \sup_{\tilde{z}_k\in(S')^n} h_1(\tilde{z}_k\given x) \prod_{i\in \mathcal{V}}\ell_2^{(i)}(Y_k^{(i)}\given \{z_k^{(i)}\}).
\end{equation}

The difficulty with \eqref{eq:hybrid_multisensor_likelihood} in a decentralised setting is that $h_1(\tilde z\given x)$ couples the sensor-specific intermediates $z^{(1)}_k,\dots,z^{(n)}_k$. We now derive independent local likelihood possibility factors that upper bound \eqref{eq:hybrid_multisensor_likelihood} and can therefore be used within the same distributed recursion as in Section~\ref{sec:decentralisedBayesianInference}. For each $i\in \mathcal{V}$, define the marginal p.f.\ of the $i$-th intermediate component by
\begin{equation}
h_1^{(i)}(z^{(i)}\given x) = \sup_{z^{(j)} \in S', j \neq i} h_1\big( [ z^{(1)},\dots,z^{(n)} ]^{\top} \given x \big), \qquad z^{(i)}\in S',\ x\in S.
\end{equation}
Let $\{w_i\}_{i\in \mathcal{V}}$ be weights such that $w_i\ge 0$ and $\sum_{i\in \mathcal{V}}w_i=1$.
Since $h_1(\tilde z\given x)\le h_1^{(i)}(z^{(i)}\given x)$ for every $\tilde{z} = [ z^{(1)},\dots,z^{(n)} ]^{\top}$ and for every $i$, we have the pointwise bound
\begin{equation}
\label{eq:h1_factorised_upper_bound}
h_1(\tilde z\given x) \leq \min_{i\in \mathcal{V}} h_1^{(i)}(z^{(i)}\given x)
\leq \prod_{i\in \mathcal{V}} h_1^{(i)}(z^{(i)}\given x)^{w_i},
\end{equation}
where the second inequality uses that $a\in[0,1]\mapsto a^{w}$ is nondecreasing and that $\min_i a_i \le \prod_i a_i^{w_i}$ for $a_i\in[0,1]$ and $\sum_i w_i=1$. Define the local hybrid likelihood at node $i$ by
\begin{equation}
\ell^{(i)}(Y_k^{(i)}\given \{x\}) = \sup_{z_k^{(i)}\in S'}
\ell_2^{(i)}(Y_k^{(i)}\given \{z_k^{(i)}\}) h_1^{(i)}(z_k^{(i)}\given x)^{w_i}.
\end{equation}
Then, using \eqref{eq:h1_factorised_upper_bound} inside \eqref{eq:hybrid_multisensor_likelihood} and the fact that, for any collection of functions $\{\psi_i\}_{i \in \mathcal{V}}$, it holds that $\sup_{\tilde z} \prod_i \psi_i(z^{(i)}) \le \prod_i \sup_{z^{(i)}}\psi_i(z^{(i)})$, we obtain
\begin{equation}
\label{eq:hybrid_independent_upper_bound}
\ell(\tilde{Y}_k\given \{x\}) \leq \prod_{i\in \mathcal{V}}\ell^{(i)}(Y_k^{(i)}\given \{x\}).
\end{equation}

Equation~\eqref{eq:hybrid_independent_upper_bound} provides an \emph{independent} collection of local likelihood possibility factors that is consistent with the original dependent mechanism in the sense that it upper bounds the multi-sensor likelihood induced by \eqref{eq:hybrid_multisensor_likelihood}. Importantly, the factorisation \eqref{eq:h1_factorised_upper_bound} is only applied to the correlated stage $h_1$, while the independent probabilistic stage $\ell_2^{(i)}$ is preserved. This yields a strictly less conservative construction than treating the entire multi-sensor likelihood as being of unknown dependence.

\subsection{Implementation}
\label{sec:implementationBernoulli}

\subsubsection{Independent observations}

For the sake of simplicity, we assume that the dynamics and likelihood are Gaussian, i.e.\ there exist matrices $G$, $Q$, $H$, $R$ of appropriate dimensions such that $g(x \given x') = \overline{\mathrm{N}}(x; Gx', Q)$ and $p(y \given x) = \mathrm{N}(y; Hx, R)$, with $Q$ and $R$ positive definite. The non-linear case has been considered in \cite{Ristic2019}. We also assume that the p.f.\ $f_{\mathrm{b}}$ is a max-mixture of Gaussian p.f.s of the form $f_{\mathrm{b}}(x) = \max_{i \in \{1, \dots, N_{\mathrm{b}}\}} w_{\mathrm{b},i} \overline{\mathrm{N}}(x; \mu_{\mathrm{b},i}, P_{\mathrm{b},i})$, with $\{w_{\mathrm{b},i}\}_{i=1}^{N_{\mathrm{b}}}$ a collection of non-negative weights such that $\max_{i \in \{1, \dots, N_{\mathrm{b}}\}} w_{\mathrm{b},i} = 1$, and with $\mu_{\mathrm{b},i}$ and  $P_{\mathrm{b},i}$ the expected value and covariance matrix of the $i$-th term, respectively. These assumptions allow to write the recursion of the possibilistic Bernoulli filter in closed form as follows. If the p.f.\ $f_{k-1}$ is a Gaussian max-mixture of the form $f_{k-1}(x) = \max_{i \in \{1, \dots, N_{k-1}\}} w_{k-1,i} \overline{\mathrm{N}}(x; \mu_{k-1,i}, P_{k-1,i})$, then the predicted p.f.\ $f_{k|k-1}$ follows as
\begin{multline*}
f_{k|k-1}(x) =  \alpha_{k|k-1}^{-1}\max\big\{ \beta_{k-1} \tau_{01} f_{\mathrm{b}}(x), \alpha_{k-1} \tau_{11} \\
\times \max_{i \in \{1, \dots, N_{k-1}\}} w_{k-1,i} \overline{\mathrm{N}}(x; G\mu_{k-1,i}, GP_{k-1,i}G^{\top} + Q)  \big\},
\end{multline*}
which is also a max-mixture of Gaussian p.f.s which can be expressed using $N_{k|k-1} = N_{k-1} + N_{\mathrm{b}}$ Gaussian terms, the $i$-th one having weight $w_{k|k-1,i}$, expected value $\mu_{k|k-1,i}$ and covariance matrix $P_{k|k-1,i}$. The expressions of $\alpha_{k|k-1}$ and $\beta_{k|k-1}$ are unaffected by the form of $f_{k-1}$ and are therefore not repeated. For the update step, the expression of $\alpha_k$ can be specialised to the Gaussian case as
\begin{multline*}
\alpha_k \propto \alpha_{k|k-1} \max_{i \in \{1,\dots,N_{k|k-1}\}} w_{k|k-1,i} \max\big\{ \beta_{\mathrm{d}} \kappa(Y_k), \\
\quad \max_{y \in Y_k} \alpha_{\mathrm{d}} \kappa(Y_k \setminus \{y\}) |2 \pi R|^{-1/2} \overline{\mathrm{N}}(x; H\mu_{k|k-1,i}, HP_{k|k-1,i}H^{\top} + R) \big\},
\end{multline*}
The updated p.f.\ $f_k$ is characterised by
\begin{multline*}
f_k(x) \propto  \max_{i \in \{1,\dots,N_{k|k-1}\}} w_{k|k-1,i} \max\big\{ \beta_{\mathrm{d}} \kappa(Y_k), \\
\max_{y \in Y_k} \alpha_{\mathrm{d}} \kappa(Y_k \setminus \{y\}) |2\pi R|^{-1/2} \overline{\mathrm{N}}(x; \tilde{\mu}_{k,i}, \tilde{P}_{k,i}) \big\}
\end{multline*}
with $\tilde{\mu}_{k,i}$ and $\tilde{P}_{k,i}$ the standard updated expected value and covariance matrix of the Kalman filter corresponding to the $i$-th predicted term. Although $f_k$ is once again a Gaussian max-mixture, it is usually necessary to apply pruning and merging to it, as is standard, in order to control the number of terms. The latter is done using the Hellinger distance due to the fact that it is a more conservative notion of distance than the Mahalanobis distance. Terms that are sufficiently close are then merged following the standard procedure, with the only difference being that the final weight of the merged term is the maximum of the weights of the terms being merged \cite{Houssineau2021linear}.

\subsubsection{Partially-unknown dependence via a shared latent term}

For the partially-unknown dependence case of Section~\ref{sec:partialDependenceHybrid}, we model the measurement generation in two stages, matching the narrative that a correlated error source exists (unknown at fusion time), while the final acquisition noise is conditionally independent:
\begin{itemize}
\item \emph{Stage 1 (correlated, possibilistic).}
At each time $k$, introduce a latent error $\bm{b}_k$ as a real vector of the same dimension as the observations, that is shared across sensors (e.g.\ common bias due to an unmodelled effect). We do not treat $\bm{b}_k$ probabilistically in the filter; instead, we describe it by a Gaussian possibility function $f_{\bm{b}_k}(b_k) = \overline{\mathrm{N}}(b_k;0,R_1)$, for some positive definite matrix $R_1$. This latent error, together with the observation function, implicitly defines the first stage possibility function $h_1(\tilde{z}_k\given x)$ as
\[
h_1(\tilde{z}_k\given x) = \sup_{b_k} \bigg( \prod_{j=1}^n \bm{1}_{\{H^{(j)}(x) + b_k\}}(z_k^{(j)}) \bigg) \overline{\mathrm{N}}(b_k;0,R_1),
\]
where $\bm{1}_{\{H^{(i)}(x) + b_k\}}$ is the indicator of $H^{(i)}(x) + b_k$, with $H^{(i)}$ the observation function of sensor $\mathrm{S}_i$.
\item \emph{Stage 2 (conditionally independent, probabilistic).}
Conditioned on $b_k$ and $x$, sensors produce independent noisy measurements
\[
y_k^{(i)} = H^{(i)}(x) + b_k + v_k^{(i)},
\qquad
v_k^{(i)} \sim \mathrm{N}(0,R_2),
\]
with $v_k^{(i)}$ and $v_k^{(j)}$, $i \neq j$, mutually independent, and $R_2$ positive definite.
\end{itemize}
Given this two-stage model, the local likelihood $p(y \given x)$, on which the definition \eqref{eq:generalLikelihood} of the Bernoulli likelihood in the independent case is based, is replaced by the mixed term
\begin{equation}
\label{eq:local-effective-likelihood}
h^{(i)}(y\given x) = \frac{1}{\sqrt{|2\pi R_2|}}\overline{\mathrm{N}}\big(y; H^{(i)}(x), R_2 + R_1/w_i\big),
\end{equation}
since the marginals of $h_1(\tilde{z}_k\given x)$ are simply $h_1^{(i)}(z^{(i)}\given x) = \overline{\mathrm{N}}(z^{(i)};H^{(i)}(x),R_1)$.
Thus, the partially-unknown dependence is handled by an \emph{inflation} of the effective measurement covariance, derived from the most informative independent upper bound induced by the shared possibilistic latent term.

\section{Decentralised Sensor fusion}
\label{sec:sensorFusion}

This section follows the approach introduced in Section~\ref{sec:decentralisedBayesianInference} and provides explicit formulae for the fusion of Gaussian max-mixture possibilistic Bernoulli filters detailed in Section~\ref{sec:BernoulliFilter}. As before, we focus on decentralised fusion for a network of $n$ sensors with nodes $\mathrm{S}_i$, $i \in \mathcal{V} = \{1,\dots,n\}$, and with connectivity represented by a graph $(\mathcal{V},\mathcal{E})$. To make the proposed approach more concrete, we describe it via the pseudo-code in Algorithms \ref{alg:decentralisedFusion}-\ref{alg:fusion}. Algorithm~\ref{alg:decentralisedFusion} describes the general recursion with prediction, update and fusion being performed sequentially at each time step. We consider these three parts in turn and justify the corresponding computations.

\begin{algorithm}[t!]
\caption{Decentralised fusion at the sensor node $\mathrm{S}_i$}
\label{alg:decentralisedFusion}
\begin{algorithmic}[1]
\Require Observation sets $(Y_k)_{k=1}^K$ at $\mathrm{S}_i$; Number of time steps $K$; Discount factor $\omega_i \in [0,1]$; Weight matrix $\Gamma$; Number of network iterations $L$
\State $N \gets 0$, $\hat{\alpha} \gets 1$, $\hat{\beta} \gets 1$
\ForAll{$k \in \{1, \dots, K\}$}
\State $(\alpha, \beta) \gets \big( \max\{ \hat{\beta} \tau_{01}^{\omega_i}, \hat{\alpha} \tau_{11}^{\omega_i} \},\, \max\{ \hat{\beta} \tau_{00}^{\omega_i}, \hat{\alpha} \tau_{10}^{\omega_i} \} \big)$
\ForAll{$j \in \{1, \dots, N\}$} \Comment{Discounted prediction}
\State $(w_j, \mu_j, P_j) \gets \big( \hat{\alpha} \tau_{11} w_j / \alpha,\, G \mu_i,\, G P_j G^{\top} + Q/\omega_i \big)$
\EndFor
\ForAll{$j \in \{1, \dots, N_{\mathrm{b}}\}$} \Comment{Birth}
\State $(w_{N+j}, \mu_{N+j}, P_{N+j}) \gets \big(\hat{\beta} \tau_{01} w_{\mathrm{b},j}^{\omega_i} / \alpha,\,  \mu_{\mathrm{b},j},\,  P_{\mathrm{b},j}/\omega_i \big)$
\EndFor
\State $N \gets N + N_{\mathrm{b}}$
\State $F^{(i)} \gets \mathrm{LocalUpdate}\big(Y_k, \alpha, \beta, (w_j, \mu_j, P_j)_{j=1}^N \big)$ \Comment{Update}
\ForAll{$l \in \{1, \dots, L\}$} \Comment{Fusion}
\State Broadcast $F^{(i)}$ and receive $\{F^{(i')} : i' \in \mathcal{N}_i  \}$ from neighbours
\ForAll{$i' \in \mathcal{N}_i$}
\State $F^{(i)} \gets \mathrm{Fusion}(F^{(i)}, F^{(i')}, \Gamma_{i,i}, \Gamma_{i,i'})$
\EndFor
\EndFor
\State $\big(\hat{\alpha}, \hat{\beta}, (w_j, \mu_j, P_j)_{j=1}^N \big) \gets F^{(i)}$ \Comment{Unpack information after fusion}
\EndFor
\end{algorithmic}
\end{algorithm}

\subsection{Prediction}

\subsubsection{Discounted Markov transition}
\label{sec:discMarkov}

As shown in Section~\ref{sec:decentralisedBayesianInference}, the Markov transition needs to be discounted to preserve the independence between sensor nodes. In the considered Gaussian max-mixture implementation of the possibilistic Bernoulli filter, the Markov transition $G(\cdot \given X_{k-1})$ on $\mathsf{X}$ is characterised by
\[
G(X_k \given \emptyset) =
\begin{cases*}
\tau_{00} & if $X_k = \emptyset$ \\
\tau_{01} f_{\mathrm{b}}(x_k) & if $X_k = \{x_k\}$,
\end{cases*}
\]
and
\[
G(X_k \given \{x_{k-1}\}) =
\begin{cases*}
\tau_{10} & if $X_k = \emptyset$ \\
\tau_{11} g(x_k \given x_{k-1}) & if $X_k = \{x_k\}$.
\end{cases*}
\]
The discounted version $G(\cdot \given X_{k-1})^{\omega}$ of this Markov transition is of the same form, with each term simply being brought to the power $\omega \in [0,1]$. This allows for \eqref{eq:BernoulliPrediction} to be used directly with, e.g., $\tau_{11}^{\omega}$ and $g(x_k \given x_{k-1})^{\omega}$ instead of $\tau_{11}$ and $g(x_k \given x_{k-1})$, respectively. Lines 3-5 in Algorithm~\ref{alg:decentralisedFusion} follow directly from these calculations and from the fact that Gaussian p.f.s are closed under exponentiation.

\begin{remark}
If $n$ is large and $\omega = 1/n$ then $g(x_k \given x_{k-1})^{\omega}$ might be extremely uninformative and make inference at each node more challenging, especially if there is a substantial number of false alarms. This can be addressed if the sensor FoVs are different, in which case the information in a given region does not have to be communicated to sensors that do not observe it. This extension is however kept for future work.
\end{remark}

\subsubsection{Discounted Gaussian max-mixture}
\label{sec:discGMM}

A related and important result that is unique to possibility theory is that Gaussian max-mixtures are closed under power, indeed, for any Gaussian max-mixture $f$, say, $f(x) = \max_{i \in \{1, \dots, N\}} w_i \overline{\mathrm{N}}(x; \mu_i, P_i)$, for some $N \in \mathbb{N}$, some non-negative weights $w_i$ such that $\max_{i \in \{1, \dots, N\}} w_i = 1$, and some collection $\{(\mu_i,P_i)\}_{i = 1}^N$ of expected values and covariance matrices, it holds that $f^{\omega}$ is also a Gaussian max-mixture for any $\omega \in [0,1]$. In particular, it holds that $f^{\omega}(x) = \max_{i \in \{1, \dots, N\}} w_i^{\omega} \overline{\mathrm{N}}(x; \mu_i, P_i/\omega)$. This result justifies the discounted birth model implemented in Line 7 of Algorithm~\ref{alg:decentralisedFusion}.

\subsection{Update}

The $\mathrm{LocalUpdate}()$ function, described in Algorithm~\ref{alg:localUpdateAtSi}, implements directly the approach of Section~\ref{sec:implementationBernoulli}. Indeed, based on the assumed conditional independence of the observations between sensor nodes, there is no redundancy to be corrected and therefore no particular discount to be applied. Line 9 of Algorithm~\ref{alg:localUpdateAtSi} (as well as Line 8 of Algorithm~\ref{alg:fusion}) is a renormalisation of the weights to ensure that they have a maximum equal to $1$. In both algorithms, this is followed by a call to a $\mathrm{Reduction}()$ function which is a Gaussian max-mixture reduction function, performing pruning and merging; the only difference with standard merging techniques is that the weight of a set of merged components is the maximum of the merged weights, as detailed in \cite{Houssineau2021linear}.

\begin{algorithm}[t!]
\caption{$\mathrm{LocalUpdate}(Y, \alpha, \beta, (w_j, \mu_j, P_j)_{j=1}^N)$}
\label{alg:localUpdateAtSi}
\begin{algorithmic}[1]
\Require Observation set $Y = \{y_1,\dots,y_M\}$; Gaussian max-mixture Bernoulli p.f.\ under the form $\big(\alpha, \beta, (w_j, \mu_j, P_j)_{j=1}^N \big)$
\ForAll{$j \in \{1, \dots, N\}$}
\State $(\hat{w}_j, \hat{\mu}_j, \hat{P}_j) \gets \big(\beta_{\mathrm{d}} \kappa(Y) w_j,\, \mu_j,\,  P_j \big)$
\Comment{Detection failure}
\ForAll{$m \in \{1, \dots, M\}$}
\Comment{Detection by $m$-th observation}
\State $j' \gets mN + j$
\State $c_{m,j} \gets |2\pi R|^{-1/2}\overline{\mathrm{N}}(y_m; H\mu_j, H P_j H^{\top} + R)$
\State $\hat{w}_{j'} \gets c_{m,j} \alpha_{\mathrm{d}} \kappa(Y \setminus \{y_m\}) w_j$
\State $(\hat{\mu}_{j'}, \hat{P}_{j'}) \gets \mathrm{KalmanUpdate}(\mu_j, P_j)$
\EndFor
\EndFor
\State $u \gets \max_j \hat{w}_j$
\State $(\hat{w}_j)_{j = 1}^{(M+1)N} \gets (\hat{w}_j)_{j = 1}^{(M+1)N} / u$
\Comment{Renormalise weights}
\State $(\tilde{w}_j, \tilde{\mu}_j, \tilde{P}_j)_{j=1}^N \gets \mathrm{Reduction}\big( (\hat{w}_j, \hat{\mu}_j, \hat{P}_j)_{j=1}^{(M+1)N} \big)$
\Comment{Pruning and merging}
\State $(\tilde{\alpha}, \tilde{\beta}) \gets \big( u\alpha, \, \kappa(Y) \beta \big)/ \max\{u\alpha, \kappa(Y) \beta\}$
\Comment{Update possibility of existence}
\State \Return $F = \big(\tilde{\alpha}, \tilde{\beta}, (\tilde{w}_j, \tilde{\mu}_j, \tilde{P}_j)_{j=1}^N \big)$
\end{algorithmic}
\end{algorithm}

\subsection{Fusion}
\label{sec:FusionImplementation}

After the local update, we have independent information at sensor node $\mathrm{S}_i$, under the form of a local posterior p.f.\ $F^{(i)}_{k}$, and aim to produce a fused p.f.\ $\hat{F}_{k}$. A direct application of \eqref{eq:fusensources} yields the parameters of $\hat{F}_{k}$ as $\hat{\beta}_k \propto \prod_{j \in \mathcal{N}_i} \beta^{(i)}_{k}$ and $\hat{\alpha}_k \propto \sup_{x \in S} \prod_{j \in \mathcal{N}_i } \alpha^{(j)}_{k} f^{(j)}_{k}(x)$ with $\max\{ \hat{\alpha}_k, \hat{\beta}_k \} = 1$, and $\hat{f}_{k}(x) \propto \prod_{j \in \mathcal{N}_i } f^{(j)}_{k}(x)$. The products in the expression of the fused p.f.\ $\hat{F}_{k}$ are computed by multiplying two terms at a time in Lines 12-13 of Algorithm~\ref{alg:decentralisedFusion}, with the specific calculations for the case of a Gaussian max-mixture being given in Algorithm~\ref{alg:fusion}. These steps are repeated $L$ times.

\begin{algorithm}[t!]
\caption{$\mathrm{Fusion}(F,F',\gamma,\gamma')$}
\label{alg:fusion}
\begin{algorithmic}[1]
\Require Two Gaussian max-mixture Bernoulli p.f.s under the form $F = \big(\alpha, \beta, (w_j, \mu_j, P_j)_{j=1}^N \big)$ and $F' = \big(\alpha', \beta', (w'_j, \mu'_j, P'_j)_{j=1}^{N'} \big)$; Discount weights $\gamma$ and $\gamma'$
\ForAll{$j \in \{1, \dots, N\}$} \Comment{Fuse terms of Gaussian max-mixture}
\ForAll{$j' \in \{1, \dots, N'\}$}
\State $m \gets (j-1) N' + j'$
\State $\hat{w}_m \gets (w_j)^{\gamma} (w'_{j'})^{\gamma'} \overline{\mathrm{N}}( \mu_j; \mu'_{j'}, P_j/\gamma +  P'_{j'}/\gamma')$
\State $\hat{P}_m \gets \big( \gamma(P_j)^{-1} +  \gamma'(P'_{j'})^{-1} \big)^{-1}$
\State $\hat{\mu}_m \gets \hat{P}_m \big(\gamma(P_j)^{-1}\mu_j +  \gamma'(P'_{j'})^{-1}\mu'_{j'} \big)$
\EndFor
\EndFor
\State $u \gets \max_j \hat{w}_j$
\State $(\hat{w}_j)_{j = 1}^{NN'} \gets (\hat{w}_j)_{j = 1}^{NN'} / u$ \Comment{Renormalise weights}
\State $(\tilde{w}_j, \tilde{\mu}_j, \tilde{P}_j)_{j=1}^N \gets \mathrm{Reduction}\big( (\hat{w}_j, \hat{\mu}_j, \hat{P}_j)_{j=1}^{NN'} \big)$ \Comment{Pruning and merging}
\State $(\tilde{\alpha}, \tilde{\beta}) \gets \big( u \alpha \alpha',\, \beta \beta' \big) / \max\{u \alpha \alpha', \beta \beta' \}$ \Comment{Fuse possibilities of existence}
\State \Return $\tilde{F} = \big(\tilde{\alpha}, \tilde{\beta}, (\tilde{w}_j, \tilde{\mu}_j, \tilde{P}_j)_{j=1}^N \big)$
\end{algorithmic}
\end{algorithm}

\subsubsection{Non-linear observations with partially-unknown dependence}

For the partially-unknown dependence case of Section~\ref{sec:partialDependenceHybrid}, the likelihood $p(y\given x)$ is replaced by $h^{(i)}(y\given x)$ as defined in \eqref{eq:local-effective-likelihood}, which is evaluated with the non-linear map $H^{(i)}$. In practice, the update step in Algorithm~\ref{alg:localUpdateAtSi} is unchanged except that Line 5 is replaced by
\[
    c_{m,j} \gets |2\pi R_2|^{-1/2}\overline{\mathrm{N}}(y_m; H^{(i)}(\mu_j), \Sigma_j^{(i)} + R_2 + R_1/w_i),
\]
where $\Sigma_j^{(i)}$ is the projection of the covariance matrix $P_j$ into the observation space via the mapping $H^{(i)}$.

\section{Simulations}
\label{sec:simulations}

In this section, we demonstrate the efficacy of the proposed approach in comparison to baselines. In Section~\ref{sec:perf_standard}, we consider a conventional tracking scenario, whereas in Section~\ref{sec:perf_weak}, we consider sensors which individually cannot observe the 2D position of the target. This scenario is particularly challenging for probabilistic representations as it requires non-informative densities over the unobserved dimensions, which is more naturally captured by possibilistic representations. We compare the performances of different methods by quantifying the error in their target number estimate, localisation, and OSPA with respect to the ground truth. We also compare the posterior uncertainties to the “oracle” centralised probabilistic baseline by their entropy. While the primary analysis assumes known parameters, we demonstrate the robustness of the proposed approach to parameter uncertainty (partially-known probability of detection) in Appendix C.

We consider simulations designed using the units of the international system, which allows for making these units implicit. Specifically, we consider $K = 25$ time steps of duration $\Delta = 1$, where targets evolve in the state space $S = \mathbb{R}^4$ according to a nearly-constant velocity model in the 2-dimensional Euclidean plane, i.e.\ $X_k = G X_{k-1} + U_k$ with $U_k \sim \mathrm{N}(0, Q)$ independently for any $k$, where
\[
G = I_2 \otimes
\begin{bmatrix}
1 & \Delta \\
0 & 1
\end{bmatrix}
\quad\text{and}\quad
Q = \sigma^2 I_2 \otimes 
\begin{bmatrix}
\Delta^4/4 & \Delta^3/2 \\
\Delta^3/2 & \Delta^2
\end{bmatrix},
\]
and $\sigma = 0.5$ is the standard deviation of the process noise. We now characterise the observation process, omitting the superscript $\cdot^{(i)}$ for parameters that are constant across sensors. We consider a linear observation model, that is $Y_k^{(i)} = H^{(i)}(X_k - x_{\mathrm{s}}^{(i)}) + V_k^{(i)}$ with $V_k^{(i)} \sim \mathrm{N}(0, \sigma'^2 I_{d'})$ independently for any $k$ and any $i$, where $\sigma' = 5$ is the standard deviation of the observational noise and where $d'$ is the dimension of the observations. We consider the case of $n = 4$ sensors, with each sensor at one of the four locations in $[500\pm300, 500\pm300]$, where any given sensor has an edge connecting it to the $2$ closest sensors. Communication between sensors is controlled by the weight matrix $\Gamma$, defined based on the \emph{Metropolis weights} \cite{Calafiore2009} as is usual in decentralised fusion. We denote by $p_{\mathrm{d}}(x) = 0.8$ the probability of detection at state $x \in S$. The number of false alarms is Poisson distributed with parameter $\lambda_{\mathrm{fa}}$, with the corresponding observations being drawn uniformly at random from the observation space for Sensor $i$, denoted $S^{\prime(i)}$.

To facilitate the interpretation of the results, the times of birth and death of the target are predetermined: it is born at time $5$ and disappears at time $20$. Yet, this information is not given to the considered filters which must rely on a much less informative birth and death model: The birth model varies between scenarios but we always consider the probability of survival at $x \in S$ to be $p_{\mathrm{s}}(x) = 1-10^{-3}$. Probabilistic and possibilistic approaches rely on pruning of the Gaussian mixture and max-mixture components with a thresholds of $5 \times 10^{-4}$ for the proposed method, $10^{-5}$ for GA, and $10^{-3}$ for AA. For the merging of components, it is based on the Hellinger distance with a threshold of $0.4$ for the proposed approach and on the Mahalanobis distance with a threshold of $8$ for the probabilistic methods.

The presence of the target is confirmed in probabilistic methods when the probability of existence is greater than a given threshold. Since arithmetic average and geometric average methods behave differently, the threshold for the former ($\tau_{\mathrm{aa}} = 0.9$) is different than the threshold for the latter ($\tau_{\mathrm{ga}} = 0.95$). The mechanisms that the proposed approach relies on are close to the ones of the geometric average fusion, i.e., products of powers of p.f.s/p.d.f.s, so that the same threshold $\tau_{\mathrm{ga}}$ is used; yet, instead of requiring the possibility of presence $\alpha$ to be greater than $\tau_{\mathrm{ga}}$, we check if the possibility of non-existence $\beta$ is smaller than $1 - \tau_{\mathrm{ga}}$. This amounts to requiring that the corresponding minimum subjective probability of presence be no smaller than $\tau_{\mathrm{ga}}$. Figure~\ref{fig:Credibility_Presence_Corr} shows the evolution of the \emph{credibility of presence}, i.e.\ either the probability of presence in the baselines or the minimum subjective probability of presence in the proposed method, for one of the considered scenarios.

An important aspect of the proposed possibilistic approach is that it does not take the above data-generating mechanism directly as a model, and instead translates it into p.f.s. The Gaussian p.f.\ describing the dynamics is assumed to take the same expected value and covariance matrices as in the true model, and target-wise parameters are translated as follows: for any $x \in S$, denoting by $\hat{S}^{(i)} = \{x \in S : H x \in S^{\prime(i)}\}$ the subset of the state space observed by Sensor $i$, we set $\alpha_{\mathrm{s}}(x)=1$, $\alpha_{\mathrm{d}}(x) = \bm{1}_{\hat{S}^{(i)}}(x)$ and $\beta_{\mathrm{d}}(x) = 1 - p_{\mathrm{d}} \bm{1}_{\hat{S}^{(i)}}(x)$, which correspond to the interpretation of p.f.\ as upper bounds for probabilities, as detailed in \cite{Houssineau2021linear}, with the upper bound being on the probability of detection failure in this case. The probabilistic methods use the true data-generating mechanism as a model.  The considered performance metrics are averaged over $1000$ Monte Carlo runs, unless otherwise stated.

For the dynamics and observation model, generating data according to a probabilistic model allows to generate varied scenarios so that a robust performance assessment can be carried out. For the target birth, although it is rare in applications that it would be a true random process, and although we actually consider a fixed birth time in experiments, we still consider a probabilistic model so that the onus is on the possibilistic method. As a consequence, the birth model must also be translated into p.f.s. In both experiments, the birth model will be a Gaussian mixture $p_{\mathrm{b}}(x) = \sum_{i=1}^{N_{\mathrm{b}}} \tilde{w}_{\mathrm{b},i} \mathrm{N}(x; \mu_{\mathrm{b},i}, P_{\mathrm{b},i})$. In the possibilistic model, we consider $f_{\mathrm{b}}(x) = \max_{i \in \{1,\dots, N_{\mathrm{b}}\}} w_{\mathrm{b},i} \overline{\mathrm{N}}(x; \mu_{\mathrm{b},i}, P_{\mathrm{b},i})$, with $w_{\mathrm{b},i} = \tilde{w}_{\mathrm{b},i} V_{\mathrm{obs}} / V_{\mathrm{b},i}$, where $V_{\mathrm{obs}} = 2\pi\sigma'$ and $V_{\mathrm{b},i} = \sqrt{|2\pi P_{\mathrm{b},i}|}$ are the ``volumes'' of the observation and of the considered birth term, respectively. The rationale for such an expression is that a large uncertainty about the state of the target at birth, relative to the observation uncertainty, will yield a small possibility of birth, as in the probabilistic case.

Additional plots regarding the two considered scenarios are given in Appendix~D.

\subsection{Standard tracking}
\label{sec:perf_standard}

\begin{figure}[htbp]
    \centering
    \begin{subfigure}[b]{0.4\textwidth}
        \centering
        \includegraphics[width=\textwidth]{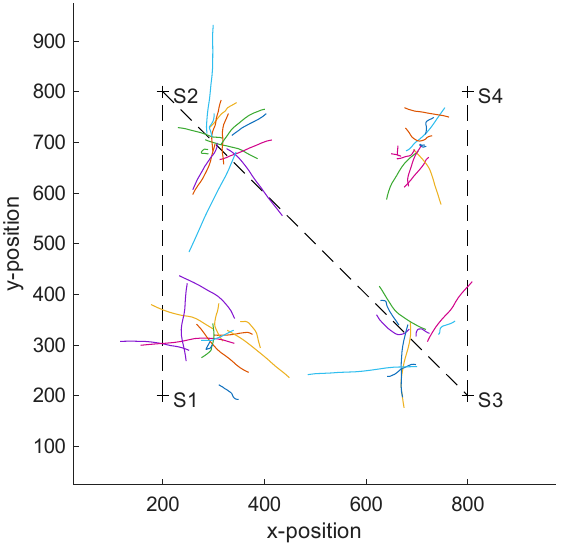}
        \caption{Standard scenario.\\\hphantom{and a new line.}}
        \label{fig:traj_realisations_Std}
    \end{subfigure}
\begin{subfigure}[b]{0.4\textwidth}
    \centering
    \includegraphics[width=\textwidth]{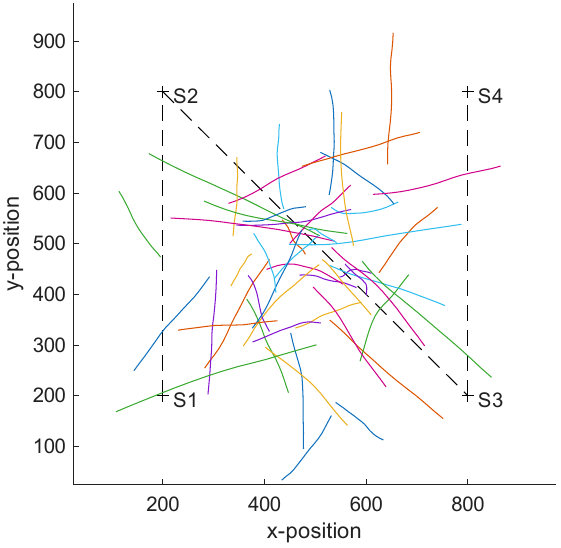}
    \caption{Scenario with weakly-informative observations.}
    \label{fig:traj_realisations_Weak}
    \end{subfigure}
    \caption{$50$ realisations of the target trajectory (colour lines), sensor locations (black $+$) and sensor links (black - -).}
    \label{fig:traj_realisations}
\end{figure}

We consider the standard linear observation such that, for any Sensor $i$, the observation matrix is
\[
H^{(i)} = 
\begin{bmatrix}
1 & 0 & 0 & 0 \\
0 & 0 & 1 & 0
\end{bmatrix},
\]
so that $d' = 2$ and the observation space is a square of size $1000$ centred on the sensor. To make this scenario more challenging, we consider an average of $\lambda_{\mathrm{fa}} = 25$ false alarms per time step. We model the birth of the target as follows: there is a probability $1/K$ for the target to be born at any given time step and, when it is born, its state is drawn from $\frac{1}{4}\sum_{i=1}^4 \mathrm{N}(\mu_{\mathrm{b},i}, \Sigma_{\mathrm{b}})$, where each of the means $\mu_{\mathrm{b},1}, \dots, \mu_{\mathrm{b},4}$ is one of $[500\pm200, 0, 500\pm200, 0]^{\top}$ and where $\Sigma_{\mathrm{b}}$ is the diagonal matrix with diagonal $[25^2, 5^2, 25^2, 5^2]$. The sensors are placed at the $4$ locations $\{500\pm300, 500\pm300\}$ and the edges on the graph are $(1,2)$, $(2,3)$, $(3,4)$, as shown in Figure~\ref{fig:traj_realisations}.

\begin{figure}
\centering
    \begin{subfigure}[b]{0.48\textwidth}
        \centering
        \includegraphics[width=\textwidth]{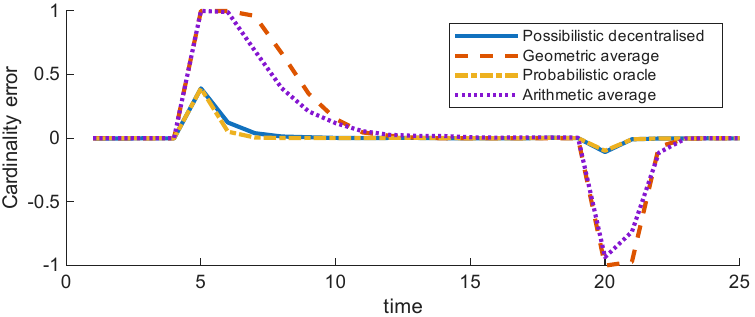}
        \caption{Cardinality error: true cardinality minus estimated cardinality}
        \label{fig:cardStd}
    \end{subfigure}
    \begin{subfigure}[b]{0.48\textwidth}
        \centering
        \includegraphics[width=\textwidth]{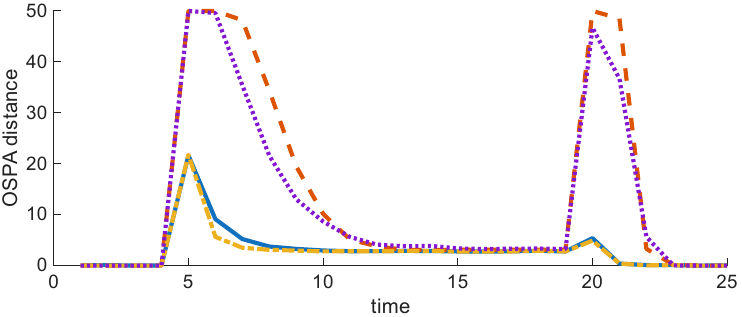}
        \caption{OSPA distance with a cut off of $50$ and with the Euclidean metric as localisation error}
        \label{fig:ospaStd}
    \end{subfigure}
    \begin{subfigure}[b]{0.48\textwidth}
        \centering
        \includegraphics[width=\textwidth]{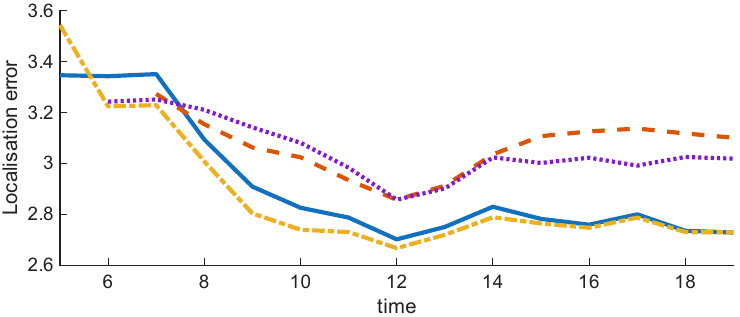}
        \caption{Localisation error based on the Euclidean distance}
        \label{fig:accStd}
    \end{subfigure}
    \begin{subfigure}[b]{0.48\textwidth}
        \centering
        \includegraphics[width=\textwidth]{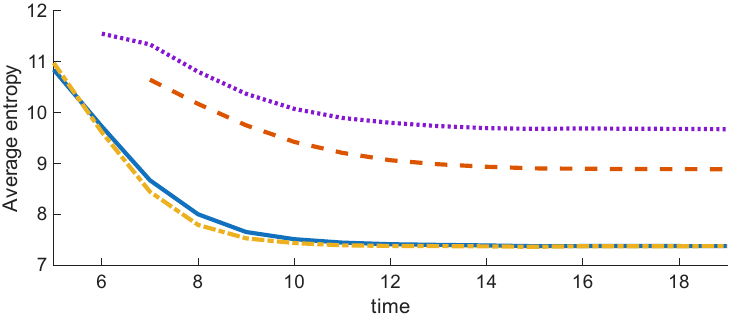}
        \caption{Entropy of the Gaussian term with highest weight}
        \label{fig:entropyStd}
    \end{subfigure}
\caption{Average performance for the standard tracking problem.}
\label{fig:4sensorStd}
\end{figure}

\figurename~\ref{fig:4sensorStd} corresponds to the case where $L=2$ iterations are performed for decentralised fusion, and shows that possibilistic fusion largely outperforms its probabilistic counterpart for several of the considered performance criteria. In \figurename~\ref{fig:ospaStd}, where the performance is assessed via the OSPA metric \cite{schuhmacher2008consistent}, it is already clear that the loss incurred by the proposed decentralised fusion algorithm is negligible when compared to the oracle. Figures~\ref{fig:cardStd} and  \ref{fig:accStd} show that the main source of performance in the proposed method is a better cardinality estimation, with similar results between all methods in terms of localisation error. The difference in cardinality error between the arithmetic average and geometric average fusion in \figurename~\ref{fig:cardStd} is in part due to the choice of confirmation threshold; it is nonetheless clear that fine tuning $\tau_{\mathrm{aa}}$ would not make the arithmetic average fusion outperform the proposed method: the OSPA distance, averaged over the $K=25$ time steps and over $100$ Monte Carlo runs, is $12.12$, $12.05$ and $12.13$ when $\tau_{\mathrm{aa}}$ equals $0.85$, $0.9$ and $0.95$, respectively.

One of the advantages of the proposed method is that the variance of the Gaussian term with highest weight is not overestimated, as opposed to arithmetic/geometric average methods. This behaviour is consistent with the fact that correlation-agnostic pooling rules (e.g., Chernoff/exponential-mixture fusion) are designed to be conservative under unknown cross-correlations \citep{gunay2017chernoff}, and related pooling strategies such as harmonic-mean density fusion have been proposed in the same spirit for mixture-based track fusion \citep{sharma2026pooling}. Accordingly, we use entropy here as a diagnostic of uncertainty inflation/conservativeness rather than as evidence of optimality. This is illustrated in \figurename~\ref{fig:entropyStd}, which shows the entropy $\log \det(2\pi e \hat{P}_k)$, with $\hat{P}_k$ the updated covariance matrix of the Gaussian term with highest weight at time $k$. Although the localisation error of the geometric average fusion is close to the proposed method, there remains a substantial error in the estimated variance, which can negatively affect downstream tasks. The localisation error and entropy are only relevant when the target exists, so \figurename~\ref{fig:accStd} and \ref{fig:entropyStd} focus on the corresponding time interval. If a track was never confirmed across the Monte Carlo runs at a given time then no results are indicated at that time.

We also study the behaviour of the proposed method and the baselines with varying number of iterations. Table~\ref{tab:AvgOSPA} gives the OSPA distance for $L = 1$, $2$ and $4$ iterations, averaged over all $K = 25$ time steps as well as over a $1000$ Monte Carlo runs. The reference, referred to as ``Asymptotic'', is obtained from a fully connected graph, in which case $L=1$ is sufficient to ensure convergence. These results further validate the proposed method by showing a convergence to the asymptotic regime, even with a limited number of iterations. Given the structure of the sensor network, a single iteration is not sufficient for the information from each sensor to reach all other sensors, and the performance is more seriously affected in this case. For reference, the averaged OSPA distance for the geometric and arithmetic average fusion with a fully connected graph are both over $10$ due to the large cardinality error.

\begin{table}[b!]
    \caption{Averaged OSPA distance for decentralised possibilistic methods over all time steps and $1000$ Monte Carlo runs.}
    \label{tab:AvgOSPA}
    \centering
    \begin{tabular}{p{30mm}|p{15mm}|p{15mm}|p{15mm}|p{15mm}}
    & $L \to \infty$ & $L=4$ & $L=2$ & $L=1$ \\
    \hline
    Possibilistic & 3.1058 & 3.1090 & 3.1096 & 3.1097 \\
    Arithmetic Average & 11.8714 & 11.9132 & 12.0550 & 12.4355 \\ 
    Geometric Average & 13.6795 & 13.7050 & 13.7929 & 13.9517
    \end{tabular}
\end{table}

The relative computational cost of the baselines compared to the proposed method in this scenario is $1.10$ for GA, and $0.58$ for AA, with GA and AA propagating approximately the same number of mixture terms (as reported in Appendix~D.1). Indeed, AA fusion is typically computationally efficient in distributed Bernoulli filtering because the fusion operator is a convex combination and does not create cross-terms \cite{li2019distributed}. 

\subsection{Tracking with weakly-informative observations}
\label{sec:perf_weak}

We consider another scenario where each sensor needs to rely more heavily on other sensors to effectively detect and track the target. In this case, each sensor observes a different component of the state: the observation matrix for Sensor~$i$ is $H^{(i)} = e_i$, with $e_i$ the unit vector with the $i$-th component equal to $1$ and the other components equal to $0$. It follows that $d'=1$ for all sensors, but the observation space differs between sensors with $S^{\prime(i)} = [H^{(i)}x^{(i)}_{\mathrm{s}}-500, H^{(i)}x^{(i)}_{\mathrm{s}}+500]$ for $i = 1,3$, i.e., when a position is observed, and $S^{\prime(i)} = [-50,50]$ for $i = 2,4$, i.e., when a velocity is observed. We consider the same sensor position and network geometry as in the previous section, as illustrated in Figure~\ref{fig:traj_realisations_Weak}. Since this scenario is already challenging due to the observation model, we consider an average of only $\lambda_{\mathrm{fa}} = 1$ false alarm per time step.

We model the birth of the target as follows: there is a probability $1/K$ for the target to be born at any given time step and, when it is born, its state is drawn from $\mathrm{N}(\mu_{\mathrm{b}}, \Sigma_{\mathrm{b}})$, with $\mu_{\mathrm{b}} = [500, 0, 500, 0]^{\top}$ and with $\Sigma_{\mathrm{b}}$ the diagonal matrix with diagonal $[100^2, 10^2, 100^2, 10^2]$. This birth model makes the state of the target at birth highly random while still ensuring that the target remains in the field of view of the sensors with high probability. Since both the birth model and the observations are weakly informative in this scenario, limiting the loss of information during the fusion is particularly important.

\begin{figure}
\centering
    \begin{subfigure}[b]{0.48\textwidth}
        \centering
        \includegraphics[width=\textwidth]{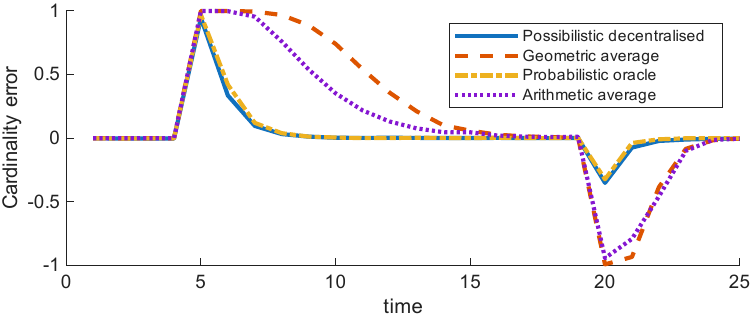}
        \caption{Cardinality error: true cardinality minus estimated cardinality}
        \label{fig:card}
    \end{subfigure}
    \begin{subfigure}[b]{0.48\textwidth}
        \centering
        \includegraphics[width=\textwidth]{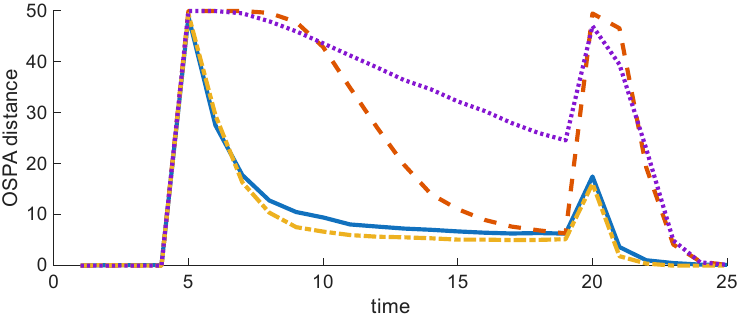}
        \caption{OSPA distance with a cut off of $50$ and with the Euclidean metric as localisation error}
        \label{fig:ospa}
    \end{subfigure}
    \begin{subfigure}[b]{0.48\textwidth}
        \centering
        \includegraphics[width=\textwidth]{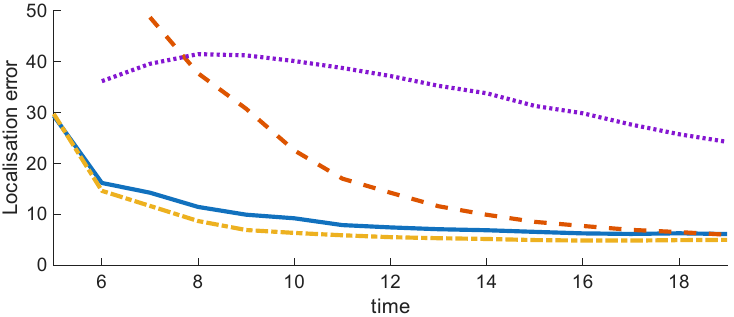}
        \caption{Localisation error based on the Euclidean distance}
        \label{fig:acc}
    \end{subfigure}
    \begin{subfigure}[b]{0.48\textwidth}
        \centering
        \includegraphics[width=\textwidth]{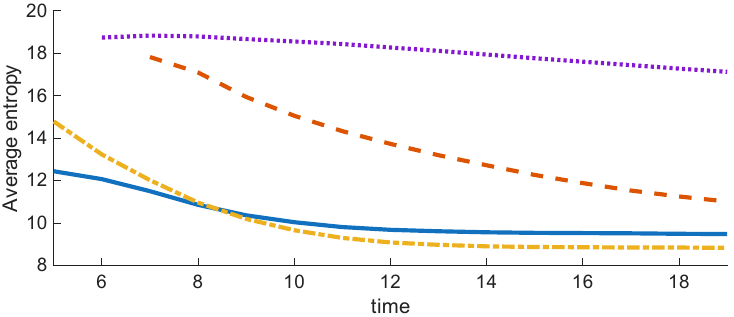}
        \caption{Entropy of the Gaussian term with highest weight}
        \label{fig:entropy}
    \end{subfigure}
\caption{Average performance over $1000$ Monte Carlo runs for the tracking problem with weakly-informative observations.}
\label{fig:4sensor}
\end{figure}

\figurename~\ref{fig:4sensor} considers the case $L=2$ and shows that, despite the difficulty of this scenario, the proposed method continues to closely align with the oracle both in terms of OSPA distance (\figurename~\ref{fig:ospa}) and average entropy (\figurename~\ref{fig:entropy}). The slightly lower entropy of the proposed method in \figurename~\ref{fig:entropy} after track confirmation can be due to more fragmented mixture terms that have yet to be merged (the plot only considers the entropy of the mixture term with highest weight). The arithmetic average fusion does not handle the considered weakly informative observation model well: Although the cardinality error decreases faster than for the geometric average fusion, as seen in \figurename~\ref{fig:card}, the localisation error remains very large, as shown in \figurename~\ref{fig:acc}, making this approach underperform in terms of OSPA distance. The slow initialisation of the geometric average fusion is likely due to the very low weight obtained when taking the product between multivariate Gaussian distributions with high variance in different components. Although the proposed approach also relies on products, the results differ due to the way Gaussian p.f.s are normalised.

\begin{figure}[htbp]
    \centering
    \begin{minipage}{0.45\textwidth}
    \centering
    \includegraphics[width=.8\textwidth]{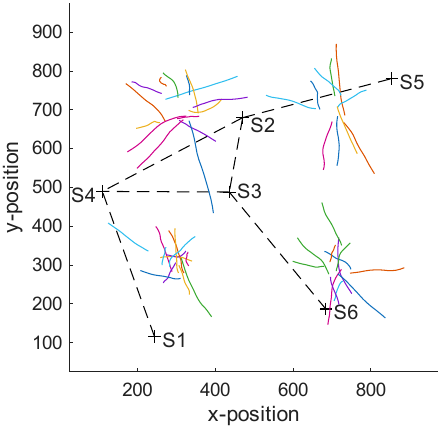}
    \caption{$50$ realisations of the target trajectory (colour lines), sensor locations (black $+$) and network links (black $--$) for the scenario with non-linear and partially-dependent observations.}
    \label{fig:traj_realisations_Corr}
    \end{minipage}
    \hfill
    \begin{minipage}{0.45\textwidth}
    \centering
    \includegraphics[width=.8\textwidth]{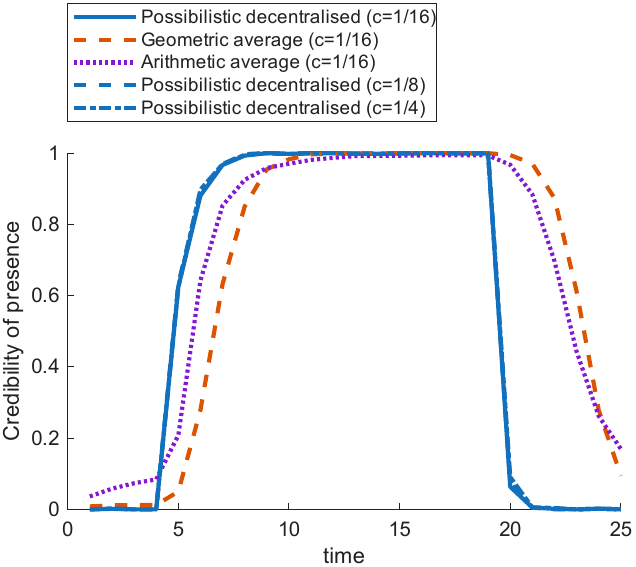}
    \caption{Credibility of presence averaged over $500$ Monte Carlo runs for the tracking problem with non-linear and partially-dependent observations.}
    \label{fig:Credibility_Presence_Corr}
    \end{minipage}
\end{figure}

\subsection{Non-linear observations with partially-unknown dependence}
\label{sec:nonlinear-partial-dependence}

e adapt the standard tracking scenario of Section~\ref{sec:perf_standard} by replacing the linear Cartesian observation model with a non-linear range--bearing model, and by introducing a partially-unknown dependence between sensors through a shared latent error term. A total of $6$ sensors are considered, with a geometry and connectivity as shown in Figure~\ref{fig:traj_realisations_Corr}. All aspects not explicitly modified below (dynamics, birth/death, false-alarm rate, sensor placement, communication graph and Metropolis weights, discounting weights, pruning/merging thresholds, etc.) are kept identical to Section~\ref{sec:perf_standard} and Appendix~D.1.

Let the target state be $x = [p_x, v_x, p_y, v_y]^\top \in S = \mathbb{R}^4$ at a given time step and let Sensor~$\mathrm{S}_i$'s position be $p_{\mathrm{s}}^{(i)}$. Define the observation function $H^{(i)}:S\to\mathbb{R}^2$ by $H^{(i)}(x) = [r^{(i)}(x), \theta^{(i)}(x) ]^{\top}$, with
\[
\delta^{(i)}(x) = \begin{bmatrix} p_x \\ p_y \end{bmatrix} - p_{\mathrm{s}}^{(i)},
\quad
r^{(i)}(x) = \|\delta^{(i)}(x)\|,
\quad
\theta^{(i)}(x) = \mathrm{atan2}\big(\delta^{(i)}_2(x),\delta^{(i)}_1(x)\big).
\]
The observation space for Sensor~$\mathrm{S}_i$ is taken as $S'^{(i)} = [0,r_{\max}] \times (-\pi,\pi]$, with $r_{\max} = 10^3$, and we define $\widehat{S}^{(i)} = \{x\in S:\ H^{(i)}(x)\in S'^{(i)}\} =\{x\in S:\ r^{(i)}(x)\le r_{\max}\}$. We consider $L=3$ iterations per time step. As in Section~\ref{sec:perf_standard}, we set $\alpha_d(x)=\bm{1}_{\hat{S}^{(i)}}(x)$ and $\beta_d(x)=1-p_d \bm{1}_{\hat{S}^{(i)}}(x)$. To define the covariance matrices $R_1$ and $R_2$, we first define a base matrix $R = \mathrm{diag}(\sigma_r^2,\sigma_{\theta}^2)$, with $\sigma_r = 5$ and $\sigma_{\theta} = \pi/180$ and then set $R_1 = c R$ and $R_2 = (1-c)R$, with $c \in \{1/16, 1/8, 1/4\}$ controlling the amount of correlated noise in the observation. The nonlinearity in the observation function is handled via an extended Kalman filter.

\begin{figure}
\centering
    \begin{subfigure}[b]{0.48\textwidth}
        \centering
        \includegraphics[width=\textwidth]{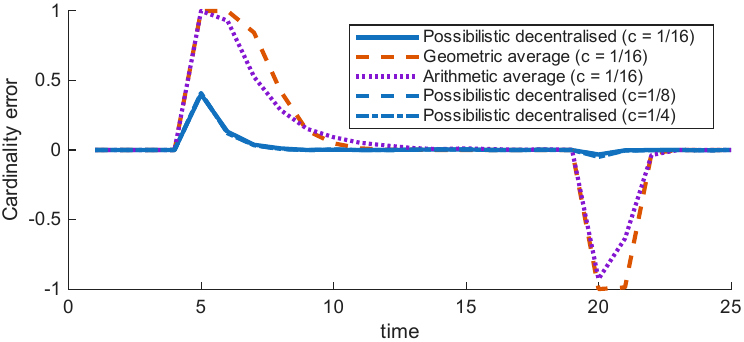}
        \caption{Cardinality error: true cardinality minus estimated cardinality}
        \label{fig:cardCorr}
    \end{subfigure}
    \begin{subfigure}[b]{0.48\textwidth}
        \centering
        \includegraphics[width=\textwidth]{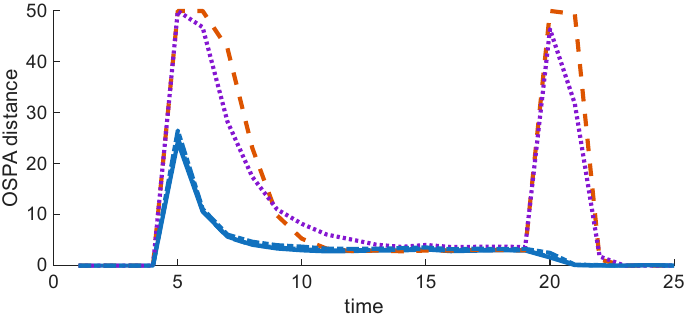}
        \caption{OSPA distance with a cut off of $50$ and with the Euclidean metric as localisation error}
        \label{fig:ospaCorr}
    \end{subfigure}
    \begin{subfigure}[b]{0.48\textwidth}
        \centering
        \includegraphics[width=\textwidth]{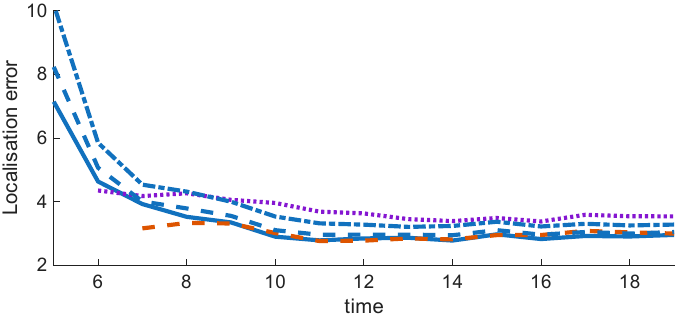}
        \caption{Localisation error based on the Euclidean distance}
        \label{fig:accCorr}
    \end{subfigure}
    \begin{subfigure}[b]{0.48\textwidth}
        \centering
        \includegraphics[width=\textwidth]{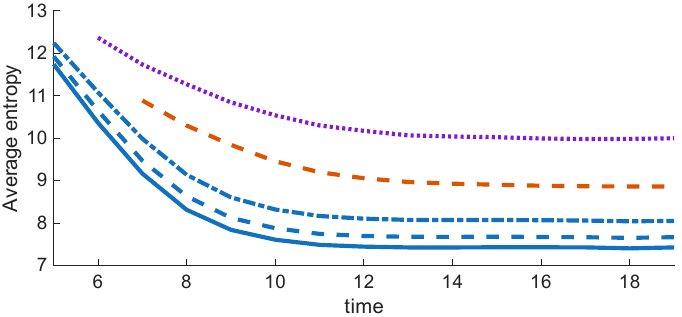}
        \caption{Entropy of the Gaussian term with highest weight}
        \label{fig:entropyCorr}
    \end{subfigure}
\caption{Average performance over $500$ Monte Carlo runs for the tracking problem with non-linear and partially-dependent observations.}
\label{fig:4sensorCorr}
\end{figure}

Figure~\ref{fig:Credibility_Presence_Corr} shows that varying the correlation parameter $c \in \{1/16, 1/8, 1/4\}$ has only a marginal effect on the proposed method: the three curves (blue) are nearly indistinguishable and the credibility of presence reaches values close to $1$ earlier than for the baselines. Figure~\ref{fig:4sensorCorr} further illustrates that this behaviour persists even when $25\%$ of the observation-noise covariance is shared across sensors (i.e.\ when $c=1/4$). For readability, the baselines are shown only for $c=1/16$: unlike the proposed method, they do not explicitly use $c$ (or an estimate thereof) in their fusion rule, and repeating the baseline curves for all values of $c$ would primarily add visual clutter. The cardinality error in Figure~\ref{fig:cardCorr} mirrors the weak dependence on $c$ observed for the credibility of presence. As $c$ increases, a larger fraction of the observation noise is common across sensors, reducing the effective amount of independent information; this is reflected by the monotone increase in entropy in Figure~\ref{fig:entropyCorr} and the corresponding degradation in localisation accuracy in Figure~\ref{fig:accCorr}. For intuition, if one estimates a common state from $n$ such correlated measurements, an effective number of independent observations is $n_{\mathrm{eff}} = n/(1+(n-1)c)$; with $n=6$ sensors and $c=1/4$, this yields $n_{\mathrm{eff}} \approx 2.67$.

\section{Conclusion}
\label{sec:conclusion}

This article introduces a decentralised inference framework based on possibility theory that is algebraically equivalent to the centralised solution. Unlike probabilistic averaging, our principled fusion rule preserves independence between sources. By applying this framework to the possibilistic Bernoulli filter, we demonstrate that it significantly outperforms geometric and arithmetic average baselines in cardinality and localisation, particularly in scenarios with weakly-informative observations. The proposed method has already been shown to generalise to sensors with limited field of view in \cite{houssineau2024decentralised}. Future work includes improving the robustness to potential calibration errors of some of the sensors, hence leveraging the capabilities of possibility theory in terms of robust inference \cite{ristic2019robust, houssineau2022robust}, as well as the extension of the approach to more challenging observation models.

\section{Acknowledgements}

J.\ Houssineau is supported by the Ministry of Education, Singapore, under its Academic Research Fund Tier 1 (RS02/24), and by the Singapore Ministry of Digital Development and Information under the AI Visiting Professorship Programme (AIVP-2024-004).

\bibliographystyle{elsarticle-num-names-alpha}
\bibliography{Uncertainty}

\input{appendix}

\end{document}

%% file: appendix.tex
\appendix

\section{Theory}
\label{app:proofs}

The results stated in this appendix are all lemmas, as opposed to results in the main article which are propositions, corollaries or theorems.

\subsection{Results in Section 3}

The following lemma focuses on splitting a single source of information and considers the general case where a p.f.\ jointly describes the information about two unknown quantities. In this general case, some information will be lost in order to obtain independence. This is an important property of p.f.s: one can trade information for statistical properties, in this case, independence. 

\begin{lemma}
\label{lemma:discountedFusion}
Let $\bm{x}$ and $\bm{y}$ be two uncertain variables on $S$ and $S'$, respectively, which are jointly described by a p.f.\ $f_{\bm{x},\bm{y}}$. Let $f_{\bm{x}}$ and $f_{\bm{y}}$  be the marginal p.f.s of $\bm{x}$ and $\bm{y}$, respectively, induced by $f_{\bm{x},\bm{y}}$ as defined in~(1). 
For any given scalar $w \in [0,1]$, let $f^{(w)}_{\bm{x},\bm{y}}$ be a function defined as
\begin{equation}
\label{eq:unknownCorrelation}
f^{(w)}_{\bm{x},\bm{y}}(x,y) = f_{\bm{x}}(x)^{w} f_{\bm{y}}(y)^{1-w},
\end{equation}
then $f^{(w)}_{\bm{x},\bm{y}}$ is a p.f.\ which verifies $f_{\bm{x},\bm{y}} \leq f^{(w)}_{\bm{x},\bm{y}}$, i.e.\ $f_{\bm{x},\bm{y}}(x,y) \leq f^{(w)}_{\bm{x},\bm{y}}(x,y)$ for all $(x,y) \in S \times S'$.
\end{lemma}

Lemma~\ref{lemma:discountedFusion} is important for pointing out that no spurious information is introduced if $f_{\bm{x},\bm{y}}$ were to be replaced by $f^{(w)}_{\bm{x},\bm{y}}$, for any $w \in [0,1]$; we will say in this kind of situation that $f^{(w)}_{\bm{x},\bm{y}}$ is \emph{consistent} with $f_{\bm{x},\bm{y}}$. This is true regardless of the strength of the correlation between $\bm{x}$ and $\bm{y}$. As a result, $f^{(w)}_{\bm{x},\bm{y}}$ can be extremely uninformative: For example, consider perfectly coupled $\bm{x}$ and $\bm{y}$ with $S = S'$ and $f_{\bm{x},\bm{y}}(x,y) = \mathbb{I}(x = y)$, where $\mathbb{I}(x = y)$ equals $1$ if $x=y$ is true and $0$ otherwise. There is then no independent information about $\bm{x}$ and $\bm{y}$ in $f_{\bm{x},\bm{y}}$ and $f^{(w)}_{\bm{x},\bm{y}} = \bm{1}$.

\begin{proof}[Proof of Lemma~\ref{lemma:discountedFusion}]
Since both marginals $f_{\bm{x}}$ and $f_{\bm{y}}$ are p.f.s, they are non-negative and have a supremum equal to 1. It follows that $f^{(w)}_{\bm{x},\bm{y}}$ is also non-negative and
\begin{align*}
\sup_{(x,y) \in S \times S'} f^{(w)}_{\bm{x},\bm{y}}(x,y) & = \sup_{(x,y) \in S \times S'} f_{\bm{x}}(x)^{w} f_{\bm{y}}(y)^{1-w} \\
& = \bigg( \sup_{x \in S} f_{\bm{x}}(x) \bigg)^w \bigg( \sup_{y \in S'} f_{\bm{y}}(y)\bigg)^{1-w} \\
& = 1,
\end{align*}
so that $f^{(w)}_{\bm{x},\bm{y}}$ is indeed a p.f. We now aim to prove that $f_{\bm{x},\bm{y}} \leq f^{(w)}_{\bm{x},\bm{y}}$. Using the definition of the joint $f^{(w)}_{\bm{x},\bm{y}}$ and of the marginals $f_{\bm{x}}$ and $f_{\bm{y}}$, we obtain
\[
f^{(w)}_{\bm{x},\bm{y}}(x,y) = 
\bigg( \sup_{y \in S'} f_{\bm{x},\bm{y}}(x,y) \bigg)^w \bigg( \sup_{x \in S} f_{\bm{x},\bm{y}}(x, y)\bigg)^{1-w}.
\]
Since it holds that
\begin{align*}
f_{\bm{x},\bm{y}}(x,y) & \leq \sup_{y' \in S'} f_{\bm{x},\bm{y}}(x,y') \\
f_{\bm{x},\bm{y}}(x,y) & \leq \sup_{x' \in S} f_{\bm{x},\bm{y}}(x',y),
\end{align*}
for any $(x,y) \in S \times S'$, we obtain that
\begin{align*}
f^{(w)}_{\bm{x},\bm{y}}(x,y) & \geq \big( f_{\bm{x},\bm{y}}(x,y) \big)^w \big( f_{\bm{x},\bm{y}}(x, y)\big)^{1-w} \\
& \geq f_{\bm{x},\bm{y}}(x,y),
\end{align*}
which concludes the proof of the proposition.
\end{proof}

Before proving Proposition 1, it is useful to introduce a general notion of conditioning for events regarding uncertain variables $\bm{u}$ and $\bm{v}$ on some sets $U$ and $V$, respectively, via
\begin{equation}
f_{\bm{u}}(u \given \bm{v} \in A ) = \dfrac{\sup_{v \in A} f_{\bm{u},\bm{v}}(u,v)}{\sup_{v \in A} f_{\bm{v}}(v)},
\label{eqn:GeneralConditional}
\end{equation}
for some subset $A$ of $V$. The possibilistic Bayes rule (3) 
corresponds to the special case where $\bm{u}=\bm{x}$, $\bm{v}=\bm{y}$, and $A = \{y\}$.

\begin{proof}[Proof of Proposition 1]
We use \eqref{eqn:GeneralConditional} with $\bm{u}=\bm{v}$, $\bm{v}=(\bm{x},\bm{z})$ and $A = \{(x,z) : x = z\} \subseteq S \times S$, and note that
$f_{\bm{u},\bm{v}} = f_{\bm{x},\bm{z},\bm{x},\bm{z}}$, which verifies
\[
f_{\bm{x},\bm{z},\bm{x},\bm{z}}(x,z,x',z') = \mathbb{I}(x=x', z=z') f_{\bm{x},\bm{z}}(x,z),
\]
so that
\[
\sup_{(x',z') : x'=z'} f_{\bm{x},\bm{z},\bm{x},\bm{z}}(x,z,x',z') = \mathbb{I}(x = z) f_{\bm{x},\bm{z}}(x,z).
\]
We then obtain the posterior p.f.\ given by
\[
f_{\bm{x},\bm{z}}(x,z \given \bm{x} = \bm{z}) = \mathbb{I}(x = z) \dfrac{f_{\bm{x},\bm{z}}(x,z)}{\sup_{x' \in S }f_{\bm{x},\bm{z}}(x',x')},
\]
for any $x,z \in S$. Note that such operations are often ill-defined for probability distributions and can lead to paradoxes (see, e.g.~\cite[Chapter 15.7]{Jaynes2003}). Since $f_{\bm{x},\bm{z}}(\cdot \given \bm{x} = \bm{z})$ takes non-zero values on the diagonal $\Delta = \{(x,x) : x \in S\}$ of $S \times S$ only, i.e.\ $f_{\bm{x},\bm{z}}(x,z \given \bm{x} = \bm{z}) = 0$ if $x \neq z$, this p.f.\ can be equivalently expressed as a univariate p.f.\ in either $\bm{x}$ or $\bm{z}$, i.e.
\begin{align*}
f_{\bm{x}}(x \given \bm{x}  = \bm{z}) & = f_{\bm{z}}(x \given \bm{x} = \bm{z}) \\ 
& = f_{\bm{x},\bm{z}}(x,x \given \bm{x} = \bm{z}),
\end{align*}
which corresponds to a change of variable by the bijective mapping $T : \Delta \to S$ defined by $T(x,x) = x$ for any~${x \in S}$.
\end{proof}

\begin{proof}[Proof of Proposition~2]
We provide a constructive proof which introduces, first, a splitting of the information in a p.f., followed by proving  the claimed equivalence.

Let us denote by $\bm{x}$ the uncertain variable described by the p.f.\ $f$ and introduce an additional uncertain variable $\bm{z}$ as a copy of $\bm{x}$. The most informative p.f.\ jointly describing $\bm{x}$ and $\bm{z}$ is
\begin{equation}
\label{eq:copying}
f_{\bm{x},\bm{z}}(x,z) = \mathbb{I}(x = z) f(x), \qquad (x,z) \in S \times S.    
\end{equation}
In turn, both $\bm{x}$ and $\bm{z}$ have $f$ as a marginal. Using \eqref{eq:unknownCorrelation}, we introduce another p.f.\ $f^{(w)}_{\bm{x},\bm{z}}$ jointly describing $\bm{x}$ and $\bm{z}$ in a way that is independent in the sense of (2), 
i.e.
\begin{equation}
\label{eq:splitting}
f^{(w)}_{\bm{x},\bm{z}}(x,z) = f(x)^{w} f(z)^{1 - w},
\end{equation}
for some $w \in [0,1]$. As shown before in Lemma~\ref{lemma:discountedFusion}, the independent description $f^{(w)}_{\bm{x},\bm{z}}$ is less informative than $f_{\bm{x},\bm{z}}$ since $f^{(w)}_{\bm{x},\bm{z}} \geq f_{\bm{x},\bm{z}}$. The two simple operations characterised by \eqref{eq:copying} and \eqref{eq:splitting} yield two independent pieces of information, modelled by the marginal p.f.s $f^w$ and $f^{1 - w}$ of $f^{(w)}_{\bm{x},\bm{z}}$, from a single piece of information expressed as $f$. 

Although $f^{(w)}_{\bm{x},\bm{z}}$ is less informative than $f_{\bm{x},\bm{z}}$, we can use the fact that $f^w$ and $f^{1-w}$ represent the same unknown quantity, i.e.\ $\bm{x} = \bm{z}$, to recover the original p.f.\ $f$ as follows. Using~\eqref{eq:splitting} in Proposition~1, 
one can fuse $f^w$ and $f^{1-w}$,~and~obtain
\begin{align*}
f_{\bm{x}}(x \given \bm{x} = \bm{z}) &= \dfrac{f^{(w)}_{\bm{x},\bm{z}}(x,x)}{\sup_{x' \in S }f^{(w)}_{\bm{x},\bm{z}}(x',x')} \nonumber \\
&= f(x),
\end{align*}
for any $x \in S$ and $w \in [0,1]$, as desired.
\end{proof}

\begin{example}
Consider the situation where $f_{\bm{x},\bm{z}}(x,z) = f_{\bm{x}}(x)$, with $f_{\bm{x}}$ the marginal p.f.\ describing $\bm{x}$, which corresponds to the absence of information about $\bm{z}$. Since $\sup_{x \in S} f_{\bm{x},\bm{z}}(x,x) = \sup_{x \in S} f_{\bm{x}}(x) = 1$, it follows that $f_{\bm{x}}(x \given \bm{x}=\bm{z}) = f_{\bm{x}}(x)$. In this case, it is clear that no p.f.\ smaller than $f_{\bm{x}}(\cdot \given \bm{x}=\bm{z})$ could correctly model the information about $\bm{x}$ since $f_{\bm{x}}$ is the only non-trivial source of information.
\end{example} 

In the context of fusion, the above result will be used together with Lemma~\ref{lemma:discountedFusion}, as detailed in the following remark.

\begin{remark}
Lemma~\ref{lemma:discountedFusion}, specifically \eqref{eq:unknownCorrelation}, provides a way to define a joint p.f.\ $f^{(w)}_{\bm{x},\bm{z}}$ when only the corresponding marginals $f_{\bm{x}}$ and $f_{\bm{z}}$ are known. Given that $\bm{x}$ and $\bm{z}$ are in the same set $S$ and using Proposition~1, 
we can fuse the obtained information given that $\bm{x} = \bm{z}$ to obtain
\[
f_{\bm{x}}( x \given \bm{x}  = \bm{z}) = \dfrac{f_{\bm{x}}(x)^w f_{\bm{z}}(x)^{1-w}}{\sup_{x' \in S} f_{\bm{x}}(x')^w f_{\bm{z}}(x')^{1-w}}.
\]
This is reminiscent of the covariance intersection technique \cite{Julier1997,Chen2002}, albeit with a different normalisation constant. In the considered context, covariance intersection follows from properties of p.f.s and from the possibilistic Bayes rule. Based on these properties, since the p.f.\ $f^{(w)}_{\bm{x},\bm{z}}$ is consistent with the underlying joint p.f.\ $f_{\bm{x,\bm{z}}}$ with $f_{\bm{x}}$ and $f_{\bm{z}}$ as marginals, for any $w \in [0,1]$, then, for any $x,z \in S$, $f_{\mathrm{min}}(x,z) = \min_{w \in [0,1]} f^{(w)}_{\bm{x},\bm{z}}(x,z)$ is also consistent with $f_{\bm{x,\bm{z}}}$. Therefore, we could also consider the posterior
\begin{equation*}
f_{\mathrm{min}}( x \given \bm{x}  = \bm{z}) = \dfrac{\min_{w \in [0,1]} f_{\bm{x}}(x)^w f_{\bm{z}}(x)^{1-w}}{\sup_{x' \in S} \min_{w \in [0,1]} f_{\bm{x}}(x')^w f_{\bm{z}}(x')^{1-w}},
\end{equation*}
where the minimisation over $w$ is performed for each $x \in S$ rather than globally.
\end{remark}

\begin{proof}[Proof of Corollary~1] 
We denote by $\{\bm{x}_i\}_{i=1}^n$ the collection of uncertain variables on $S$ independently described by the p.f.s $\{f_i\}_{i=1}^n$, then one can define the joint p.f.\ $f_{\bm{x}_{1:n}}(x_1,\dots,x_n) = \prod_{i=1}^n f_i(x_i)$ using the independence property. Following the same steps as in the proof of Proposition~1, 
the fusion of these $n$ sources of information is found to be
\begin{align*}
    \hat{f}(x) & = f_{\bm{x}_{1:n}}(x,\dots,x \given \bm{x}_i = \bm{x}_j,\, i,j \in \{1,\dots,n\}) \\
    & = \dfrac{\prod_{i=1}^n f_i(x)}{\sup_{x' \in S}\prod_{i=1}^n f_i(x')},
\end{align*}
as required.
\end{proof}

\subsection{Results in Section 4}

\begin{proof}[Proof of Corollary~1]
We prove this result by induction. Suppose without loss of generality that $\mathcal{V} = \{1,\dots,n\}$ and assume that $f^{(i,l-1)}$ and $f^{(i',l-1)}$ are independent for any $i \neq i'$. Following the same approach as in the proof of Proposition~2, 
the p.f.\ $f^{(i,l-1)}$ can be first split into $n$ independent p.f.s as
\[
\tilde{f}^{(i,l-1)}(x_{i,1},\dots,x_{i,n}) = \prod_{j\in\mathcal{V}} \Big( f^{(i,l-1)}(x_{i,j})\Big)^{\Gamma_{i,j}}.
\]
Then, leveraging the independence between $\tilde{f}^{(i,l-1)}$ and $\tilde{f}^{(i',l-1)}$ for all $i \neq i'$, the joint p.f.\ describing the information at all sensors can be written as $\prod_{i \in \mathcal{V}} \tilde{f}^{(i,l-1)}(x_{i,1},\dots,x_{i,n})$, which can be expressed as
\[
\prod_{i \in\mathcal{V}} \tilde{f}^{(i,l-1)}(x_{i,1},\dots,x_{i,n}) = \prod_{j \in \mathcal{V}} \bigg( \underbrace{\prod_{i \in \mathcal{V} } \Big( f^{(i,l-1)}(x_{i,j})\Big)^{\Gamma_{i,j}}}_{\doteq \tilde{f}^{(j,l)}(x_{1,j},\dots,x_{n,j})} \bigg).
\]
Finally, Corollary~1 
allows to conclude that the p.f.\ $f^{(j,l)}$, characterised by $f^{(j,l)}(x) \propto \tilde{f}^{(j,l)}(x,\dots,x)$, is independent of $f^{(j',l)}$ for any $j \neq j'$. Since the p.f.s are independent when $l=0$, the base case is also verified.
\end{proof}

\begin{proof}[Proof of Theorem~1]
Let us express the asymptotic posterior $f^{(i,\infty)}$ in a more explicit fashion~as
\begin{align*}
    f^{(i,\infty)}(x) & \propto \prod_{i \in \mathcal{V}} \Big[ h^{(i)}(y_i \given x) \sup_{x' \in S} g^{\varpi_i}(x \given x') f^{\omega_i}(x') \Big]^{\Gamma^*_{i,j}} \\
    & = \prod_{i \in \mathcal{V}} h^{(i)}(y_i \given x)^{\Gamma^*_{i,j}} \sup_{x' \in S} g^{\varpi_i \Gamma^*_{i,j}}(x \given x') \big( f(x') \big)^{\omega_i \Gamma^*_{i,j}}.
\end{align*}
\paragraph{Sufficient conditions}
If Assumptions A1 and A2 
both hold, the expression above becomes
\begin{align*}
    f^{(i,\infty)}(x) &  \propto \prod_{i \in \mathcal{V}} h^{(i)}(y_i \given x)^{1/n} \sup_{x' \in S} g^{\omega_i/n}(x \given x') \big( f(x') \big)^{\omega_i/n} \\
    & = \Big[\prod_{i \in \mathcal{V}} h^{(i)}(y_i \given x) \prod_{i \in \mathcal{V}} \big(\sup_{x' \in S} g(x \given x') f(x')\big)^{\omega_i}\Big]^{1/n} \\
    & = \Big[\prod_{i \in \mathcal{V}} h^{(i)}(y_i \given x) \sup_{x' \in S} g(x \given x') f(x')\Big]^{1/n}.
\end{align*}
Using (7) 
concludes.
\paragraph{Necessary conditions}
For the method to be asymptotically exact for any form of likelihood function $h^{(i)}(y_i \given x)$ at some arbitrary $x \in S$, then by (7) 
the weights $\Gamma^*_{i,j}$ must be such that 
\[
\prod_{i \in \mathcal{V}} \alpha_i^{\Gamma^*_{i,j}} = \prod_{i \in \mathcal{V}} \alpha_i^{1/n}
\]
for all $\alpha_i \in [0,1]$, which holds if and only if $\sum_{i \in \mathcal{V}}(\Gamma^*_{i,j} - 1/n) \log \alpha_i = 0$, that is, if and only if $\Gamma^*_{i,j} = 1/n$ for all $i \in \mathcal{V}$.

We can now write $(f^{(i,\infty)}(x))^n \propto \tilde{f}(x) \prod_{i \in \mathcal{V}} h^{(i)}(y_i \given x)$ with
\[
\tilde{f}(x) = \prod_{i \in \mathcal{V}} \sup_{x' \in S} g^{\varpi_i }(x \given x') f^{\omega_i }(x').
\]
For the method to be asymptotically exact for any number of sensors and nodes, it must hold in particular for the following case: Consider the $n$-sensor case where $f$ equals $\overline{\mathrm{N}}(0, \tau^{-1})$ and $g(\cdot \given x')$ equals $\overline{\mathrm{N}}(x', \tau'^{-1})$. The predicted p.f.\ at node $i \in \mathcal{V}$ is $\overline{\mathrm{N}}(0, 1/(\tau'\varpi_i) + 1/(\tau\omega_i))$. It follows that $\tilde{f}$ is $\overline{\mathrm{N}}(0, \hat{\tau}^{-1})$ with precision
\[
\hat{\tau} = \sum_{i \in \mathcal{V}}\frac{\varpi_i \omega_i \tau \tau'}{\varpi_i\tau' + \omega_i\tau}.
\]
Since the predictive p.f.\ in (7) 
simplifies to $\overline{\mathrm{N}}(0, 1/\tau + 1/\tau')$ in that case, it must hold that $\hat{\tau} = \tau\tau' / (\tau + \tau')$. Since the equality must hold for all $\tau$ and all $\tau'$, we aim to obtain a polynomial with $\tau$ and $\tau'$ as variables so that the coefficients can be matched. The polynomial in question is
\[
\tau\tau'\prod_{j \in \mathcal{V}} ( \omega_j \tau + \varpi_j \tau' ) = (\tau + \tau') \sum_{i \in \mathcal{V}} \varpi_i \omega_i \tau \tau' \prod_{j \neq i} ( \omega_j \tau + \varpi_j \tau' ).
\]
The terms of highest degrees are of the form $\tau' \tau^{n+1}$ and $\tau \tau'^{n+1}$. By matching the coefficients of the former, we obtain $\prod_{j \in \mathcal{V}} \omega_j = \sum_{i \in \mathcal{V}} \varpi_i \omega_i \prod_{j \neq i} \omega_j$, which simplifies to $\sum_{i \in \mathcal{V}} \varpi_i = 1$. By symmetry, we also obtain $\sum_{i \in \mathcal{V}} \omega_i = 1$ by matching coefficients of $\tau \tau'^{n+1}$. Now considering coefficients of terms of the form $\tau'^2\tau^n$, we obtain
\begin{align}
& \sum_{i \in \mathcal{V}} \varpi_i \prod_{j \neq i} \omega_j = \sum_{i \in \mathcal{V}} \varpi_i \omega_i \sum_{k \neq i} \varpi_k \prod_{j \neq i,k} \omega_j + \sum_{i \in \mathcal{V}} \varpi_i \prod_{j \in \mathcal{V}} \omega_j \nonumber \\
\label{eq:coeffInTauPrime2TauN}
& \iff \sum_{i \in \mathcal{V}} \frac{\varpi_i}{\omega_i} = \sum_{i \in \mathcal{V}} \varpi_i \sum_{k \neq i} \frac{\varpi_k}{\omega_k} + 1.
\end{align}
The first term on the right hand side can be re-expressed as
\begin{align*}
\sum_{i \in \mathcal{V}} \varpi_i\sum_{k \neq i} \frac{\varpi_k}{\omega_k} & = \sum_{i \in \mathcal{V}} \varpi_i \bigg( \sum_{k \neq i} \frac{\varpi_k}{\omega_k} + \frac{\varpi_i}{\omega_i} - \frac{\varpi_i}{\omega_i} \bigg) \\
& = \sum_{i \in \mathcal{V}} \frac{\varpi_i}{\omega_i} - \sum_{i \in \mathcal{V}} \frac{\varpi_i^2}{\omega_i}.
\end{align*}
Plugging this back into \eqref{eq:coeffInTauPrime2TauN}, we obtain $\sum_{i \in \mathcal{V}} \frac{\varpi_i^2}{\omega_i} = 1$. To conclude we find that
\[
\sum_{i \in \mathcal{V}} \frac{(\varpi_i - \omega_i)^2}{\omega_i} = \sum_{i \in \mathcal{V}} \frac{\varpi_i^2}{\omega_i} - 2\sum_{i \in \mathcal{V}}\varpi_i + \sum_{i \in \mathcal{V}} \omega_i = 0.
\]
It follows that $(\varpi_i - \omega_i)^2 = 0$ for all $i \in \mathcal{V}$, that is $\varpi_i = \omega_i$ for all $i \in \mathcal{V}$, as required.
\end{proof}

\section{Likelihood for the possibilistic Bernoulli filter}
\label{sec:LikelihoodForThePossibilisticBernoulliFilter}

We first consider separately the probabilistic and possibilistic components of the likelihood. We model the uncertainty about the detection of the target as epistemic, with a p.f.\ characterised by $\alpha_{\mathrm{d}}$ for a detection and $\beta_{\mathrm{d}}$ for a detection failure. The uncertainty in the data association is considered epistemic as there is a true data association which is simply unknown. For the sake of simplicity, we re-express the observation set $Y$ as a vector $\tilde{y} = (y_1,\dots,y_m)$. Assuming that a detection has occurred, a possible association is defined as choosing a component of $\tilde{y}$ as the true detection, so that the other components are all false alarms. There is no prior information about the data association, which we model as permutations of indices in $\{1,\dots,m\}$, so we choose $\bm{1}$ as the corresponding p.f.

If the target exists, is at state $x \in S$, and is detected, then the true observation $y$ is characterised by the conditional p.d.f.\ $p(\cdot \given x)$ and the false alarms are characterised by $\kappa$, with an underlying cardinality distribution $c$ on $\mathbb{N}$ and, assuming there are $n$ false alarms, with a p.d.f.\ $p_n$ on $(S')^n$. In case of detection, the uncertainty in the observation is defined via an \emph{outer probability measure} $\bar{P}_{\mathrm{d}}$ \cite{Houssineau2018parameter}, characterised by
\begin{multline}
\label{eq:obsOPM}
\bar{P}_{\mathrm{d}}(\varphi \given x) = \sum_{n \geq 1} c(n-1) \max_{\varsigma} \int \varphi(y_1,\dots,y_n) \\ \times p(y_{\varsigma(1)} \given x) p_{n-1}(y_{\varsigma(2)},\dots,y_{\varsigma(n)})  \mathrm{d} y_1 \dots \mathrm{d} y_n
\end{multline}
for any real-valued function $\varphi$ taking vectors of observations as argument, where the maximum ranges over all permutations $\varsigma$ of $\{1,\dots,n\}$. The outer probability measure $\bar{P}_{\mathrm{d}}(\cdot \given x)$ follows from the operations
\begin{enumerate}
    \item The number of false alarms is drawn from $c$,
    \item Given that there are $n-1$ false alarms, all the observation indices are shuffled in a deterministic way via a permutation $\varsigma$ from $\{1,\dots,n\}$ to itself.
    \item All observations are drawn from $p(y_{\varsigma(1)} \given x) p_{n-1}(y_{\varsigma(2)},\dots,y_{\varsigma(n)})$, 
\end{enumerate}
The order of the operations is important in this case since placing the maximum before the integral or before the sum in \eqref{eq:obsOPM} would result in a different outer probability measure. Setting $\varphi = \delta_{(y_1,\dots,y_m)}$ in $\bar{P}_{\mathrm{d}}(\varphi \given x)$ yields
\begin{multline*}
    \bar{P}_{\mathrm{d}}(y_1,\dots,y_m \given x) = c(m-1) \\
    \times \max_{\varsigma} p(y_{\varsigma^{-1}(1)} \given x) p_{m-1}(y_{\varsigma^{-1}(2)},\dots,y_{\varsigma^{-1}(m)}),
\end{multline*}
which can be simplified using the definition of $p_{n-1}$ as
\[
\bar{P}_{\mathrm{d}}(y_1,\dots,y_m \given x) = \max_{y \in Y} p(y \given x) \kappa(Y \setminus \{y\}).
\]
Following the same steps, we find that the likelihood for the case where the target is not detected is simply $\kappa(Y)$. Finally, taking the epistemic uncertainty regarding detection into account, we obtain the likelihood of interest, i.e.
\begin{equation*}
\ell(Y \given \{x\}) = \max\big\{ \beta_{\mathrm{d}} \kappa(Y), \max_{y \in Y} \alpha_{\mathrm{d}} p(y \given x) \kappa(Y \setminus \{y\}) \big\}.
\end{equation*}

\section{Robustness to parameter uncertainty}
\label{app:robustness}

In this section, we provide additional simulation results to demonstrate the robustness of the proposed approach when precise model parameters are only partially available. This addresses the capability of possibility theory to handle epistemic uncertainty regarding sensor characteristics.

We rely on the standard tracking scenario introduced in Section 7.1, with the observation model modified as follows:
\begin{enumerate*}[label=\roman*)]
\item for each detection, the true probability of detection is sampled uniformly from the interval $[0.7, 0.9]$ and $p_{\mathrm{d}}$ is set to $0.7$ in the algorithms and,
\item the standard deviation of the observational noise is set to $\sigma' = 10$ to make the scenario more challenging.
\end{enumerate*}

\begin{figure}[htbp]
\centering
    \begin{subfigure}[b]{0.48\textwidth}
        \centering
        \includegraphics[width=\textwidth]{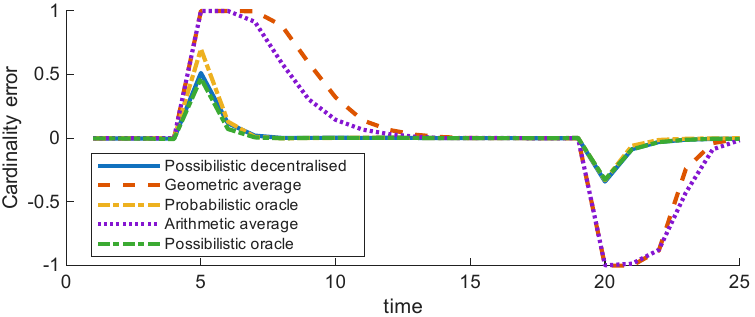}
        \caption{Cardinality error: true cardinality minus estimated cardinality}
        \label{fig:cardInt}
    \end{subfigure}
    \begin{subfigure}[b]{0.48\textwidth}
        \centering
        \includegraphics[width=\textwidth]{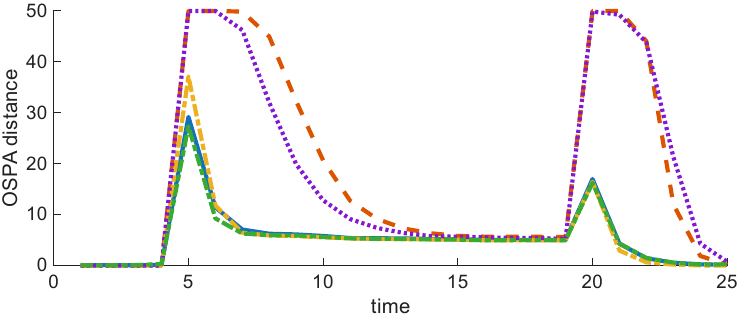}
        \caption{OSPA distance with a cut off of $50$ and with the Euclidean metric as localisation error}
        \label{fig:ospaInt}
    \end{subfigure}
    \begin{subfigure}[b]{0.48\textwidth}
        \centering
        \includegraphics[width=\textwidth]{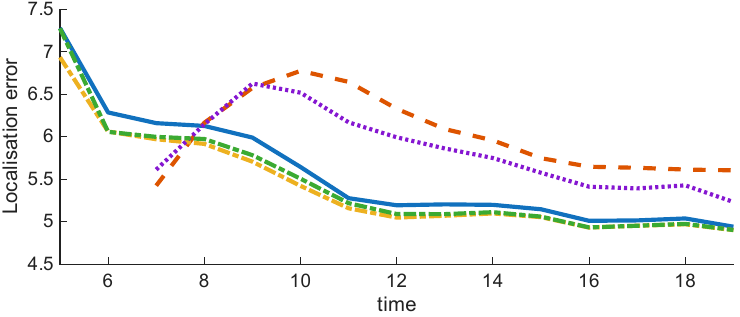}
        \caption{Localisation error based on the Euclidean distance}
        \label{fig:accInt}
    \end{subfigure}
    \begin{subfigure}[b]{0.48\textwidth}
        \centering
        \includegraphics[width=\textwidth]{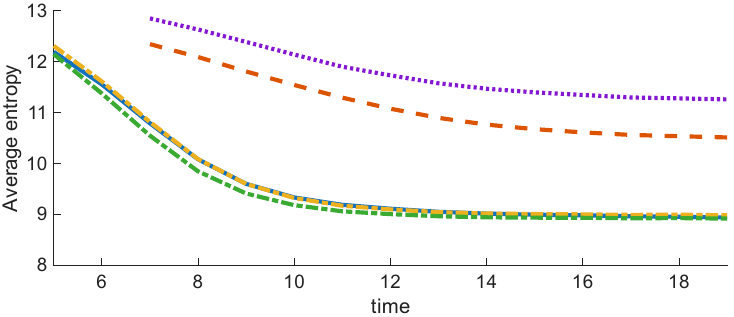}
        \caption{Entropy of the Gaussian term with highest weight}
        \label{fig:entropyInt}
    \end{subfigure}
\caption{Average performance over $1000$ Monte Carlo runs for the tracking problem with interval probability of detection.}
\label{fig:perf_int}
\end{figure}

Figure~\ref{fig:perf_int} displays the performance metric for this scenario, averaged over 1000 Monte Carlo runs. The possibilistic oracle is also shown in Figure~\ref{fig:perf_int} and displays slightly better performance compared to the probabilistic oracle due to its more flexible assumption on the probability of detection. The proposed decentralised possibilistic approach performs almost exactly as well as the possibilistic oracle, and slightly better than the probabilistic oracle at the level of the target initialisation. This demonstrates that some of the previously-identified advantages of possibilistic inference, e.g., in \cite{Ristic2019}, continue to be relevant. Yet, in this case, the vast majority of the gains against the baselines (AA \& GA) come from the information fusion capabilities of possibility theory.

\section{Additional information on Section 7}

\subsection{Standard tracking}

\begin{figure}[htbp]
    \centering
    \includegraphics[width=.5\textwidth]{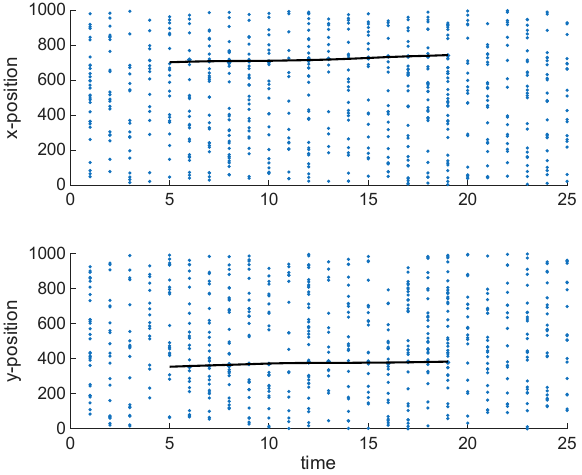}
    \caption{All observations (blue dots) and true trajectory (black line)  for the standard scenario.}
    \label{fig:XandYvsT}
\end{figure}

For this scenario, the main difficulty is in the density of false alarms, which is illustrated for one of the 4 sensors in Figure~\ref{fig:XandYvsT}. In probabilistic decentralised tracking, the loss of information in the fusion rules means that false alarms are more difficult to filter out.

\begin{figure}[htbp]
    \centering
    \includegraphics[width=.5\textwidth]{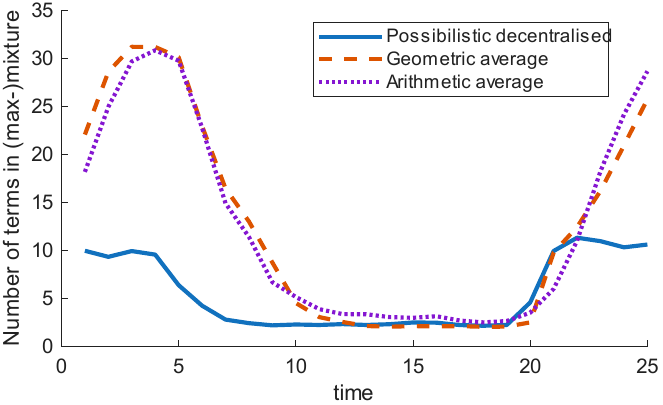}
    \caption{Average number of (max-)Gaussian-mixture terms at each communication iteration, averaged over $100$ Monte Carlo runs.}
    \label{fig:nofTErms}
\end{figure}

In addition, the evolution of the number of terms in the propagated (max)-mixtures of Gaussians is displayed in Figure~\ref{fig:nofTErms}. The number of terms in GA and AA is similar due to the choice of pruning threshold for each method ($10^{-5}$ and $10^{-3}$ respectively). The reduced number of terms in the proposed method could be due to the fact that the associated Gaussian terms have typically lower uncertainty (as confirmed by the entropy in Figure~3d), hence reducing the number of cross-terms in the fusion step.

\subsection{Weakly-informative observations}

\begin{figure}[htbp]
    \centering
    \begin{subfigure}[b]{0.4\textwidth}
        \centering
        \includegraphics[width=\textwidth]{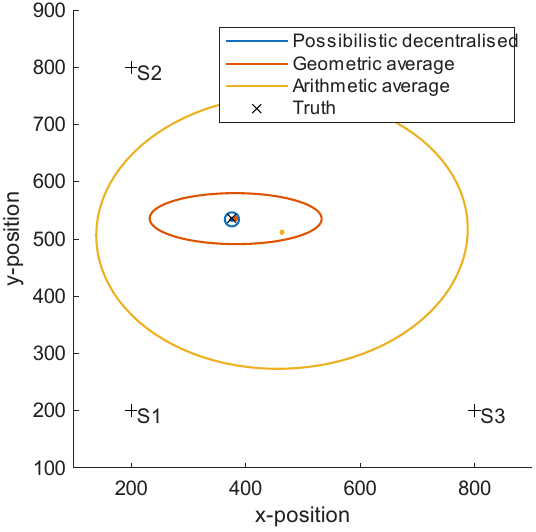}
        \caption{Position}
        \label{fig:var_pos}
    \end{subfigure}
    \hspace{3em}
    \begin{subfigure}[b]{0.4\textwidth}
        \centering
        \includegraphics[width=\linewidth]{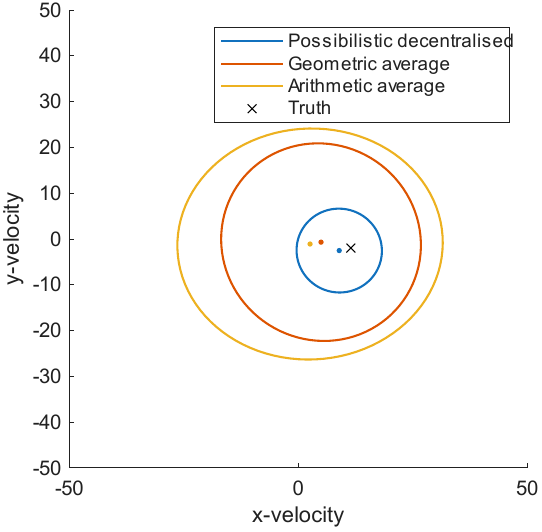}
        \caption{Velocity}
        \label{fig:var_vel}
    \end{subfigure}
    \caption{99\% ellipses for the hypothesis with largest weight at time step $7$ for the different methods in the weakly-informative observations.}
    \label{fig:var_pos_vel}
\end{figure}

To illustrate the difficulty of this scenario, Figure~\ref{fig:var_pos_vel} displays the 99\% ellipses for the posterior variance of the hypothesis with highest weight for each decentralised method at a specific time step. A time step of $7$ is considered since it emphasises the difference between the proposed approach and the baselines: the former makes optimal use of the information in the observations, which allows to quickly reduce uncertainties; by finding a consensus instead, GA and AA lose a large amount of information, which causes the posterior uncertainty to be overestimated, sometimes by orders of magnitude.

